\documentclass[fleqn,usenatbib]{mnras}

\usepackage[T1]{fontenc}
\usepackage{ae,aecompl}

\usepackage{times} 
\usepackage{tikz}
\usetikzlibrary{positioning}
\usepackage{xspace,ulem}

\usetikzlibrary{backgrounds}
\usetikzlibrary{arrows}
\usetikzlibrary{calc}
\usepackage{booktabs}
\usepackage{graphicx}	
\usepackage{amsmath}	
\usepackage{amssymb}	
\usepackage{times}
\usepackage{newtxtext}
\usepackage[varvw]{newtxmath}

\title[SZ effect in modified gravity]{The impact of modified gravity on the Sunyaev-Zel'dovich effect}

\author[M. A. Mitchell et al.]{
Myles A. Mitchell,$^{1}$\thanks{E-mail: m.a.mitchell@durham.ac.uk}
Christian Arnold,$^{1}$
C\'esar Hern\'andez-Aguayo$^{2,3,1}$
and Baojiu Li$^{1}$
\\
$^{1}$Institute for Computational Cosmology, Department of Physics, Durham University, South Road, Durham DH1 3LE, UK\\
$^{2}$Max-Planck-Institut f\"ur Astrophysik, Karl-Schwarzschild-Str. 1, D-85748, Garching, Germany\\
$^{3}$Excellence Cluster ORIGINS, Boltzmannstrasse 2, D-85748 Garching, Germany\\
}

\date{Accepted XXX. Received YYY; in original form ZZZ}

\pubyear{2019}

\begin{document}
\label{firstpage}
\pagerange{\pageref{firstpage}--\pageref{lastpage}}
\maketitle

\begin{abstract}
We study the effects of two popular modified gravity theories, which incorporate very different screening mechanisms, on the angular power spectra of the thermal (tSZ) and kinematic (kSZ) components of the Sunyaev-Zel'dovich effect. Using the first cosmological simulations that simultaneously incorporate both screened modified gravity and a complete galaxy formation model, we find that the tSZ and kSZ power spectra are significantly enhanced by the strengthened gravitational forces in Hu-Sawicki $f(R)$ gravity and the normal-branch Dvali-Gabadadze-Porrati model. Employing a combination of non-radiative and full-physics simulations, we find that the extra baryonic physics present in the latter acts to suppress the tSZ power on angular scales $l\gtrsim3000$ and the kSZ power on all tested scales, and this is found to have a substantial effect on the model differences. Our results indicate that the tSZ and kSZ power can be used as powerful probes of gravity on large scales, using data from current and upcoming surveys, provided sufficient work is conducted to understand the sensitivity of the constraints to baryonic processes that are currently not fully understood.
\end{abstract}




\section{Introduction}
\label{sec:introduction}

The Sunyaev-Zel'dovich (SZ) effect is caused by the inverse Compton scattering of cosmic microwave background (CMB) photons off of high-energy electrons within ionised gas. The effect is made up of two measurable components: a thermal (tSZ) component which arises due to the random thermal motions of the electrons; and a (much smaller) kinematic (kSZ) component resulting from the bulk motion of the gas relative to the CMB rest frame \citep[e.g.,][]{1972CoASP...4..173S,1980ARA&A..18..537S}. The tSZ and kSZ signals are both highly correlated with the presence of large-scale structures such as groups and clusters of galaxies. Their power spectra are therefore extremely sensitive to the values of cosmological parameters which affect the growth of large-scale structure, offering the possibility of probing a wide range of cosmological models, including modified gravity (MG) theories in which the strength of gravity is enhanced.

A number of works have made use of the tSZ power spectrum to constrain cosmological parameters including the dimensionless matter density $\Omega_{\rm M}$, the linear density fluctuation $\sigma_8$, the dark energy equation of state parameter and the neutrino mass \citep[e.g.,][]{Horowitz:2016dwk,Hurier:2017jgi,Bolliet:2017lha,Salvati:2017rsn}. Meanwhile, as the precision of measurements of the kSZ power continues to improve, a number of works have identified this as a promising probe for future constraints of dark energy and MG theories \citep[e.g.,][]{Ma:2013taq,Bianchini:2015iaa,Roncarelli:2018kud}. The wealth of high-quality observational data coming from current and upcoming surveys \citep[e.g.,][]{Sievers:2013ica,Aghanim:2015eva,George:2014oba,Reichardt:2020jrr,Ade:2018sbj,Abazajian:2016yjj} for both the tSZ and kSZ effects make it an exciting time for this growing area. 

In addition to their sensitivity to cosmology, the tSZ and kSZ power spectra are also highly sensitive to non-gravitational processes, such as star formation, cooling and stellar and black hole feedback, which can alter the thermal state of the intra-cluster medium \citep[e.g.,][]{McCarthy:2013qva,Park:2017amo}. Without a careful consideration of these processes, which are still not fully understood, this could pose a barrier to making reliable constraints. In recent years, a lot of progress has been made in developing sub-resolution models that allow these baryonic processes to be incorporated in cosmological simulations \citep[e.g.,][]{Schaye:2014tpa,2017MNRAS.465.3291W,Pillepich:2017jle}. For example, the IllustrisTNG simulations \citep[see, e.g.,][]{Nelson:2017cxy,Springel:2017tpz,Marinacci:2017wew,Pillepich:2017fcc,2018MNRAS.477.1206N} feature a calibrated model \citep{2017MNRAS.465.3291W,Pillepich:2017jle} that is able to produce a galaxy population whose stellar and gas properties closely match observations. Incorporating these `full-physics' models in numerical simulations, along with the cosmological model of interest, is now a vital step in order to understand the potential sensitivity of the constraints to baryonic physics.

Over the past two decades, a wide range of MG theories \citep[see, e.g.,][]{Koyama:2015vza} have been developed in an attempt to explain the late-time accelerated expansion of the Universe. These theories often predict additional forces which can alter the strength of gravity on large scales and leave observational signatures. For example, the abundance of galaxy clusters can be altered relative to the standard $\Lambda$-cold-dark-matter ($\Lambda$CDM) paradigm. Two popular MG theories include the $f(R)$ gravity \citep[see, e.g.,][]{Sotiriou:2008rp,DeFelice:2010aj} and Dvali-Gabadadze-Porrati \citep[DGP,][]{DVALI2000208} models. These predict the presence of an additional `fifth force' which enhances the strength of gravity and, in order to comply with the very tight constraints of gravity within the solar system \citep{Will:2014kxa}, both models feature screening mechanisms that can suppress the fifth force where necessary.

The $f(R)$ gravity model is representative of a wide class of theories that employ chameleon screening. Previous studies have made use of various large-scale structure probes including the halo mass function \citep[e.g.,][]{PhysRevD.92.044009,Liu:2016xes,Peirone:2016wca}, the cluster gas mass fraction \citep[e.g.,][]{Li:2015rva}, redshift-space distortions \citep[e.g.,][]{Bose:2017dtl,2018NatAs...2..967H,Hernandez-Aguayo:2018oxg}, the SZ profile \citep{DeMartino:2013zua,deMartino:2016xso} and the mass-temperature relation \citep{Hammami:2016npf} of clusters, the clustering of clusters \citep{Arnalte-Mur:2016alq}, and weak lensing by voids \citep{Cautun:2017tkc}, to test this class of models. Other works have constrained $f(R)$ gravity by comparing weak lensing measurements with X-ray and SZ observations of clusters \citep[e.g.,][]{Terukina:2013eqa,Wilcox:2015kna}.

The DGP model consists of two branches: a \textit{self-accelerating} (sDGP) branch and a \textit{normal} (nDGP) branch. The latter is often preferred since it is able to reproduce the late-time cosmic acceleration without suffering from the ghost instabilities that exist in the former. One caveat is that nDGP requires a small amount a dark energy in order to be viable; however, it is nevertheless a useful toy model that is representative of a wide class of models that exhibit Vainshtein screening \citep{VAINSHTEIN1972393}. Previous works have studied and constrained nDGP using probes including redshift-space distortions \citep[e.g.,][]{PhysRevD.94.084022,Hernandez-Aguayo:2018oxg} and cosmic voids \citep[e.g.,][]{Falck:2017rvl,10.1093/mnras/stz022}. Tests of other gravity models that also feature the Vainshtein mechanism include a comparison of weak lensing measurements of the Coma Cluster with X-ray and SZ observations \citep[][]{Terukina:2015jua}.

In this work, we study the effects of $f(R)$ gravity and nDGP on the tSZ and kSZ power spectra. These are expected to be altered by the effects of the fifth force on the abundance and peculiar motion of large-scale structures, and on the temperature of the intra-cluster gas via the enhancement of the halo gravitational potential \citep[e.g.,][]{He:2015mva,Mitchell:2020aep}. We make use of the first cosmological simulations that simultaneously incorporate galaxy formation (full-physics\footnote{We note that `full-physics' refers to the most advanced baryonic models that are currently implemented in our cosmological simulations, rather than a complete description of the underlying physics.}) plus $f(R)$ gravity \citep{Arnold:2019vpg} and nDGP \citep{Hernandez-Aguayo:2020kgq}. We measure the power spectra using mock maps of the tSZ and kSZ effect, which are generated using the simulation data. The full-physics simulations employ the IllustrisTNG galaxy formation model and we also study non-radiative simulations (using the same cosmological parameters and initial conditions), allowing us to single out fifth force and baryonic feedback effects.

The paper is structured as follows: in Sec.~\ref{sec:theory}, we briefly describe the $f(R)$ gravity and nDGP models; then, in Sec.~\ref{sec:methods}, we introduce the simulations used in this work and our methods for predicting the SZ power spectra; our main results are presented in Sec.~\ref{sec:results}; and, finally, we present a summary of our findings and their significance in Sec.~\ref{sec:conclusions}. 

Throughout the paper, an over-bar (e.g., $\bar{x}$) is used to denote the mean background value of a quantity and a subscript $_0$ indicates a present-day value, unless otherwise stated. Greek indices are used to label spacetime and run over $0,1,2,3$.

\section{Theory}
\label{sec:theory}

In this work, we have studied two models: $f(R)$ gravity and nDGP. These are described in Secs.~\ref{sec:theory:fR} and \ref{sec:theory:nDGP}, respectively. Throughout this section, we use the unit convention $c=1$, where $c$ is the speed of light.

\subsection{\boldmath $\lowercase{f}(R)$ gravity}
\label{sec:theory:fR}

The $f(R)$ gravity model is an extension of 
General Relativity (GR). It is constructed by adding a non-linear function, $f(R)$, of the Ricci scalar curvature, $R$, to the Einstein-Hilbert action of GR:
\begin{equation}
    S=\int {\rm d}^4x\sqrt{-g}\left[\frac{R+f(R)}{16\pi G}+\mathcal{L}_{\rm M}\right],
\label{eq:fR_action}
\end{equation}
where $g$ is the determinant of the metric tensor $g_{\alpha\beta}$, $G$ Newton's constant and $\mathcal{L}_{\rm M}$ 
the Lagrangian density for matter. We only consider non-relativistic matter in this paper, as we are interested in the late-time behaviour.

Taking the variation of this action with respect to the metric $g_{\alpha\beta}$ yields the modified Einstein field equations, which now include an extra tensor, $X_{\alpha \beta}$:
\begin{equation}
    G_{\alpha \beta} + X_{\alpha \beta} = 8\pi GT_{\alpha \beta},
\label{eq:modified_field_equations}
\end{equation}
where
\begin{equation}
    X_{\alpha \beta} = f_RR_{\alpha \beta} - \left(\frac{f}{2}-\Box f_R\right)g_{\alpha \beta} - \nabla_{\alpha}\nabla_{\beta}f_R.
\label{eq:GR_modification}
\end{equation}
The tensors $G_{\alpha \beta}$, $R_{\alpha \beta}$ and $T_{\alpha \beta}$ represent the Einstein tensor, the Ricci tensor and the stress-energy tensor, respectively. $\nabla_{\alpha}$ is the covariant derivative associated with the metric, and $\Box$ denotes the d'Alembert operator. The derivative $f_R\equiv{\rm d}f(R)/{\rm d}R$ represents an extra scalar degree of freedom, which can be treated as an additional scalar field whose dynamics is governed by the trace of the modified Einstein equations,
\begin{equation}
    \Box f_R = \frac{1}{3}(R - f_RR + 2f + 8\pi G\rho_{\rm M}),
\end{equation}
where $\rho_{\rm M}$ is the matter density. The scalar field mediates the fifth force of the theory, which is able to act on scales smaller than the Compton wavelength:
\begin{equation}
    \lambda_{\rm C} = a^{-1}\left(3\frac{{\rm d}f_R}{{\rm d}R}\right)^{\frac{1}{2}},
\label{eq:compton_wavelength}
\end{equation}
where $a$ is the cosmic scale factor. 

The fifth force is an attractive force felt by massive particles. In low-density environments, it enhances the strength of gravity by a factor of $1/3$. However, in high-density regions the fifth force is suppressed and GR is recovered. This is caused by the chameleon screening mechanism, which was included in the $f(R)$ model \citep[e.g.,][]{Khoury:2003aq,Khoury:2003rn,Mota:2006fz} to avoid conflict with the tight constraints from solar system tests \citep{Will:2014kxa}. The chameleon screening is brought about by an environment-dependent effective mass, $m_{\phi}=\lambda_{\rm C}^{-1}$, of the scalar field that becomes very heavy in dense regions.

In the weak-field and quasi-static limits, the modified Poisson equation, which governs structure formation in the $f(R)$ model, is given by \citep[see, e.g.,][]{Li:2011vk,Capozziello:2012ie}:
\begin{equation}
    \nabla^2\Phi = \frac{16\pi G}{3}\delta\rho_{\rm M} - \frac{1}{6}\delta R,
\end{equation}
where $\Phi$ is the Newtonian gravitational potential and $\delta\rho_{\rm M}$ and $\delta R$ represent the perturbations of the matter density and Ricci scalar, respectively. The scalar field, $f_R$, satisfies:
\begin{equation}
    \nabla^2f_R = \frac{1}{3}(\delta R - 8\pi G\delta\rho_{\rm M}).
\end{equation}

In this work, we examine the Hu-Sawicki (HS) model \citep{Hu:2007nk}, which is a popular variant of $f(R)$ gravity that is able to explain the late-time acceleration while also showing consistency with solar system tests. The model uses the following prescription for the function $f(R)$:
\begin{equation}
    f(R) = -m^2\frac{c_1\left(-R/m^2\right)^n}{c_2\left(-R/m^2\right)^n+1},
\label{eq:hu_sawicki}
\end{equation}
where $m^2\equiv8\pi G\bar{\rho}_{\rm M,0}/3=H_0^2\Omega_{\rm M}$, with $H_0$ being the Hubble constant, $\bar{\rho}_{\rm M,0}$ the present-day background matter density and $\Omega_{\rm M}$ the present-day dimensionless matter density parameter. The theory has three model parameters: $c_1$, $c_2$ and $n$. By choosing $n=1$, which is commonly used in cosmological simulations of HS $f(R)$ gravity, and making the assumption that $-R\gg m^2$, it then follows that:
\begin{equation}
    f_R \simeq -\frac{c_1}{c_2^2}\left(\frac{m^2}{-R}\right)^2.
\label{eq:fR}
\end{equation}
We also assume that the model has a background expansion history that is practically indistinguishable from that of $\Lambda$CDM, in which case the background curvature, $\bar{R}$, is given by:
\begin{equation}
    \bar{R} \simeq -3m^2\left[a^{-3} + 4\frac{\Omega_{\Lambda}}{\Omega_{\rm M}}\right],
\label{eq:ricci_scalar}
\end{equation}
where $\Omega_{\Lambda}=1-\Omega_{\rm M}$. For a realistic choice of cosmological parameters, we find that $-\bar{R}\gg m^2$ indeed holds, and so Eq.~(\ref{eq:fR}) holds for background values, such that:
\begin{equation}
    \bar{f}_R(a) = \bar{f}_{R0}\left(\frac{\bar{R}_0}{\bar{R}(a)}\right)^2.
\label{eq:fR_background}
\end{equation}
In this simplified form of the HS model, we are able to work with just a single parameter: $\bar{f}_{R0}$, the present-day background scalar field (we will omit the over-bar of $\bar{f}_{R0}$ for the remainder of this work). A higher value of $|f_{R0}|$ corresponds to a stronger modification of GR, allowing regions of higher density to be unscreened at a given time. In this work, we examine HS $f(R)$ gravity with $|f_{R0}|=10^{-6}$ and $|f_{R0}|=10^{-5}$ and refer to these models as F6 and F5, respectively.

\subsection{The nDGP model}
\label{sec:theory:nDGP}

The nDGP model assumes that the Universe is a 4-dimensional brane which is embedded in a 5-dimensional bulk spacetime. The model has an action that consists of two terms, with one being the usual Einstein-Hilbert action of GR and the other being the equivalent of the Einstein-Hilbert action, as extended to the 5 dimensions of the bulk:
\begin{equation}
    S=\int_{\rm brane} {\rm d}^4x\sqrt{-g}\left(\frac{R}{16\pi G}\right) + \int {\rm d}^5x\sqrt{-g^{(5)}}\left(\frac{R^{(5)}}{16\pi G^{(5)}}\right),
\label{eq:DGP_action}
\end{equation}
where $g^{(5)}$, $R^{(5)}$ and $G^{(5)}$ are the equivalents of $g$, $R$ and $G$ in the bulk. A characteristic scale can be defined, known as the cross-over scale $r_{\rm c}$: 
\begin{equation}
    r_{\rm c} = \frac{1}{2}\frac{G^{(5)}}{G},
\label{eq:DGP_crossover}
\end{equation}
which represents the length scale at which the behaviour of gravity transitions from 4D to 5D. The second term of Eq.~(\ref{eq:DGP_action}) will dominate on scales larger than the cross-over scale, and gravity becomes 5D. Assuming a homogeneous and isotropic background, the variation of Eq.~(\ref{eq:DGP_action}) leads to the modified Friedmann equation:
\begin{equation}
    \frac{H(a)}{H_0} = \sqrt{\Omega_{\rm M}a^{-3} + \Omega_{\rm DE}(a) + \Omega_{\rm rc}} - \sqrt{\Omega_{\rm rc}},
\label{eq:DGP_friedmann}
\end{equation}
where we have added a dark energy component to this model given that the nDGP model on its own cannot predict a late-time accelerated expansion. $\Omega_{\rm DE}(a)$ is the density parameter for this additional component, and we have assumed that $\Omega_{\rm DE}(a)$ takes such a form as to make $H(a)$ in Eq.~\eqref{eq:DGP_friedmann} identical to a $\Lambda$CDM background history. We have also assumed that the dark energy component has negligible clustering on the sub-horizon scales that we are interested in here. $\Omega_{\rm rc}$ is given by:
\begin{equation}
    \Omega_{\rm rc} \equiv \frac{1}{4H_0^2r_{\rm c}^2}.
\label{eq:DGP_omega}
\end{equation}
Deviations from $\Lambda$CDM due to the fifth force are often quantified by $H_0r_{\rm c}$ for which, according to Eq.~(\ref{eq:DGP_friedmann}), a larger value represents a smaller departure from GR. In this work, we analyse models with $H_0r_{\rm c}=5$ and $1$, and refer to these as N5 and N1, respectively. 

In the weak-field and quasi-static limits, the modified Poisson equation in the nDGP model is given by \citep{PhysRevD.75.084040}:
\begin{equation}
    \nabla^2\Phi = 4\pi Ga^2\delta\rho_{\rm M} + \frac{1}{2}\nabla^2\varphi,
\label{eq:DGP_pot}
\end{equation}
The extra scalar field of the model, $\varphi$, satisfies the following dynamical equation of motion:
\begin{equation}
    \nabla^2\varphi + \frac{r_{\rm c}^2}{3\beta a^2}\left[(\nabla^2\varphi) - (\nabla_i\nabla_j\varphi)(\nabla^i\nabla^j\varphi)\right] = \frac{8\pi Ga^2}{2\beta}\delta\rho_{\rm M},
\label{eq:DGP_scalar_field}
\end{equation}
where the function $\beta$ is given by:
\begin{equation}
    \beta(a) = 1 + 2Hr_{\rm c}\left(1 + \frac{\dot{H}}{3H^2}\right) = 1 + \frac{\Omega_{\rm M}a^{-3} + 2\Omega_{\Lambda}}{2\sqrt{\Omega_{\rm rc}\left(\Omega_{\rm M}a^{-3} + \Omega_{\Lambda}\right)}}.
\label{eq:DGP_beta}
\end{equation}
On sufficiently large scales, the non-linear terms in the square bracket of Eq.~(\ref{eq:DGP_scalar_field}) can be ignored and gravity is enhanced by a factor of $[1 + 1/(3\beta)]$. Since $\beta$ is always decreasing with time, the force of gravity is stronger at later times and has present-day enhancements of approximately $1.12$ for N1 and $1.04$ for N5. On small scales, the nonlinearity of the scalar field can no longer be ignored, causing the fifth force to be suppressed via the Vainshtein screening mechanism \citep{VAINSHTEIN1972393}.

\section{Simulations and methods}
\label{sec:methods}

In Sec.~\ref{sec:methods:simulations}, we describe the simulations used in this work. Then, in Sec.~\ref{sec:methods:maps}, we present our procedure for generating SZ maps from the simulation data.

\subsection{Simulations}
\label{sec:methods:simulations}

The results discussed in this work have been produced using the {\sc shybone} simulations \citep{Arnold:2019vpg,Hernandez-Aguayo:2020kgq}. These were run using the \textsc{arepo} code \citep{2010MNRAS.401..791S}, and they employ the IllustrisTNG galaxy formation model \citep{2017MNRAS.465.3291W,Pillepich:2017jle}. The IllustrisTNG model includes (sub-resolution) prescriptions for a number of physical processes which are necessary to reproduce a realistic galaxy population in hydrodynamical cosmological simulations: it is built using a magneto-hydrodynamics solver on \textsc{arepo}'s moving Voronoi mesh with Powell $\nabla\cdot B$ cleaning, and a magnetic field which is seeded with $1.6\times10^{-10}~\mathrm{Gauss}$ at $z=127$ \citep[see][for details]{Pakmor2011, Pakmor2013}. The model also accounts for black hole evolution and feedback (black holes are seeded in friends-of-friends (FOF) haloes above a certain mass), where the accretion is Eddington-limited Bondi-Hoyle accretion and the feedback is either thermal in the black hole proximity or a black-hole-driven kinetic wind, depending on the black hole's accretion state \citep{2017MNRAS.465.3291W, Vogelsberger2013}. Along with gas cooling and UV heating the TNG model also includes star formation employing a Chabrier initial mass function. As the stars evolve, the chemical enrichment of the gas around them is kept track of. Galactic winds from star-forming gas are expelled isotropically and are gas-metallicity-dependent \citep[see][for details]{Pillepich:2017jle}.

In addition to these full-physics simulations, we also ran non-radiative counterparts for the $f(R)$ model, using identical initial conditions and cosmological parameters. We have not run non-radiative counterparts for the nDGP model since these are computationally expensive to perform and there is already a lot of information provided by the existing simulations.

The simulations have been run for the HS $f(R)$ gravity and nDGP models using an MG solver which has been implemented in the \textsc{arepo} code. Both models feature a highly non-linear scalar field which is computed on the adaptively refining mesh (AMR grid) of \textsc{arepo}'s MG solver (see \citealt{Arnold:2019vpg, Hernandez-Aguayo:2020kgq}, for details). Once the scalar field is computed, the force field is computed on the grid for both models and the forces are interpolated from the grid to the particles using an inverse cloud-in-cell scheme. The adaptive time-stepping scheme of the code makes use of the fact that the MG forces are suppressed in high-density regions by screening mechanisms, allowing the computationally very expensive fifth force calculation to be performed less frequently than the standard gravity / hydro computations in these regions. This makes the code highly-efficient and allows the large number of high-resolution simulations which we examine in this paper to be run.

The simulations were carried out in a cubic box of a comoving length $62h^{-1}{\rm Mpc}$, employing $512^3$ dark matter particles and the same number of initial gas cells, with a mass resolution of $m_{\rm DM}=1.28\times10^8h^{-1}M_{\odot}$ and $m_{\rm gas}\approx2.5\times10^7h^{-1}M_{\odot}$, respectively. The runs start at $z=127$ with the same initial conditions in each gravity model. Particle data has been saved at various snapshots: the $f(R)$ data consists of 46 snapshots between $z=3$ and $z=0$, whereas the nDGP data includes 100 snapshots between $z=20$ and $z=0$.

The runs all use the same background cosmology: ($h$, $\Omega_{\rm M}$, $\Omega_{\rm B}$, $\Omega_{\Lambda}$, $n_{\rm s}$, $\sigma_8$) $=$ ($0.6774$, $0.3089$, $0.0486$, $0.6911$, $0.9667$, $0.8159$), where $\Omega_{\rm B}$ is the dimensionless baryonic matter density parameter, $h=H_0/(100\, {\rm km \, s^{-1}Mpc^{-1}})$, $n_{\rm s}$ is the power-law index of the primordial matter power spectrum and $\sigma_8$ is the root-mean-square of the linear matter density fluctuations over the scale of $8h^{-1}{\rm Mpc}$ at $z=0$. The $f(R)$ runs include data for F6 and F5 and the nDGP runs include data for N5 and N1. We also ran standard gravity ({GR / $\Lambda$CDM}) simulations using identical initial conditions for comparison.

In calculating the gas temperature, we assume constant values $X_{\rm H}=0.76$ and $\gamma=5/3$ for the hydrogen mass fraction and the adiabatic index, respectively. For the non-radiative data, we assume that the gas is composed entirely of ionised hydrogen and helium.

Haloes have been identified from the particle data using the \textsc{subfind} code \citep{springel2001} implemented in {\sc arepo}, which deploys the FOF algorithm and gravitational un-binding to identify groups and sub-haloes. In this work, we define the halo mass, $M_{500}$, as the total mass within a sphere (centred on the minimum of the gravitational potential) that encloses an average density which is 500 times the critical density of the Universe.

\begin{figure*}
\centering
\includegraphics[width=0.93\textwidth]{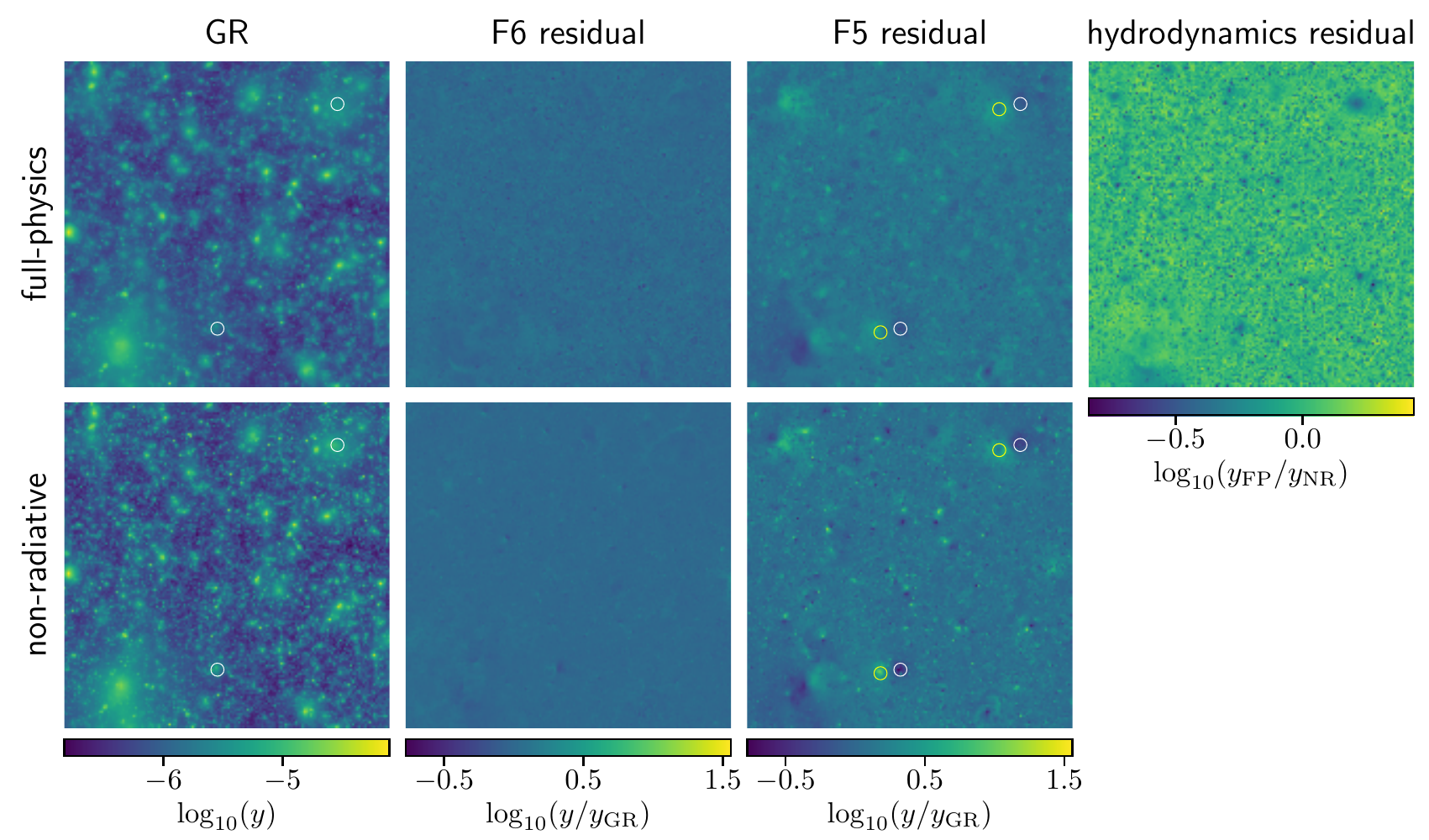}
\caption{(Colour Online) Maps of the thermal SZ effect in GR (\textit{first column}), its relative enhancement in F6 (\textit{second column}) and F5 (\textit{third column}) with respect to GR, and the relative difference between the full-physics and non-radiative GR maps (\textit{fourth column}). The maps have a side length of $1^\circ$ and a $512\times512$-pixel resolution, and have been constructed from the {\sc shybone} simulations (see Sec.~\ref{sec:methods}). Both the full-physics (\textit{top row}) and non-radiative (\textit{bottom row}) runs are shown. The $y$-parameter is computed for each pixel using Eq.~(\ref{eq:pixel_ysz}). The rings indicate two haloes whose positions are shifted in F5 (\textit{yellow}) relative to GR (\textit{white}).}
\label{fig:thermal_sz}
\end{figure*}

\begin{figure*}
\centering
\includegraphics[width=0.93\textwidth]{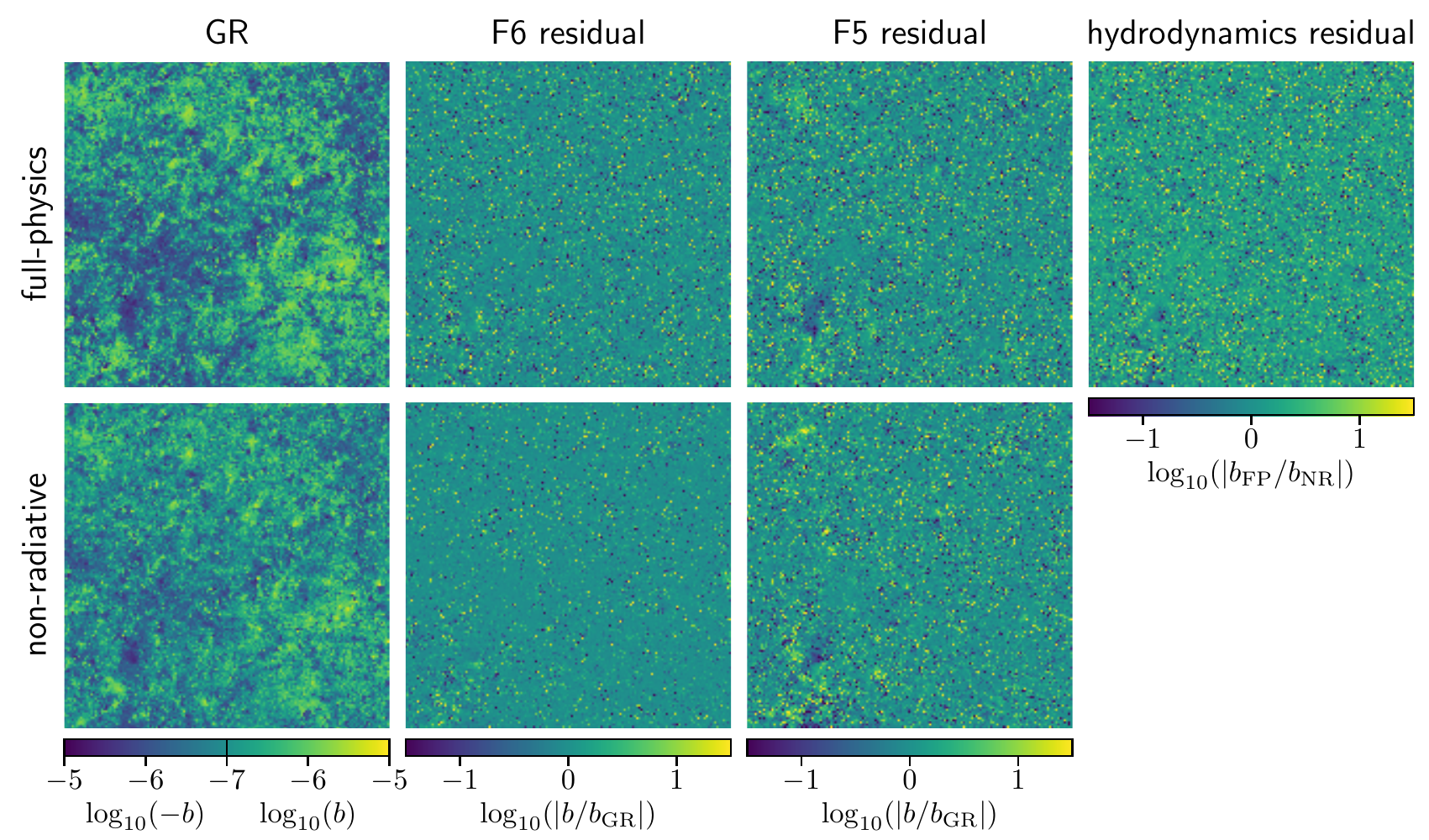}
\caption{(Colour Online) Maps of the kinetic SZ effect in GR (\textit{first column}), its absolute relative enhancement in F6 (\textit{second column}) and F5 (\textit{third column}) with respect to GR, and the absolute relative difference between the full-physics and non-radiative GR maps (\textit{fourth column}). The maps have a side length of $1^\circ$ and a $512\times512$-pixel resolution, and have been constructed from the {\sc shybone} simulations (see Sec.~\ref{sec:methods}). Both the full-physics (\textit{top row}) and non-radiative (\textit{bottom row}) runs are shown. The $b$-parameter is computed for each pixel using Eq.~(\ref{eq:pixel_bsz}).}
\label{fig:kinetic_sz}
\end{figure*}

\begin{figure*}
\centering
\includegraphics[width=0.7\textwidth]{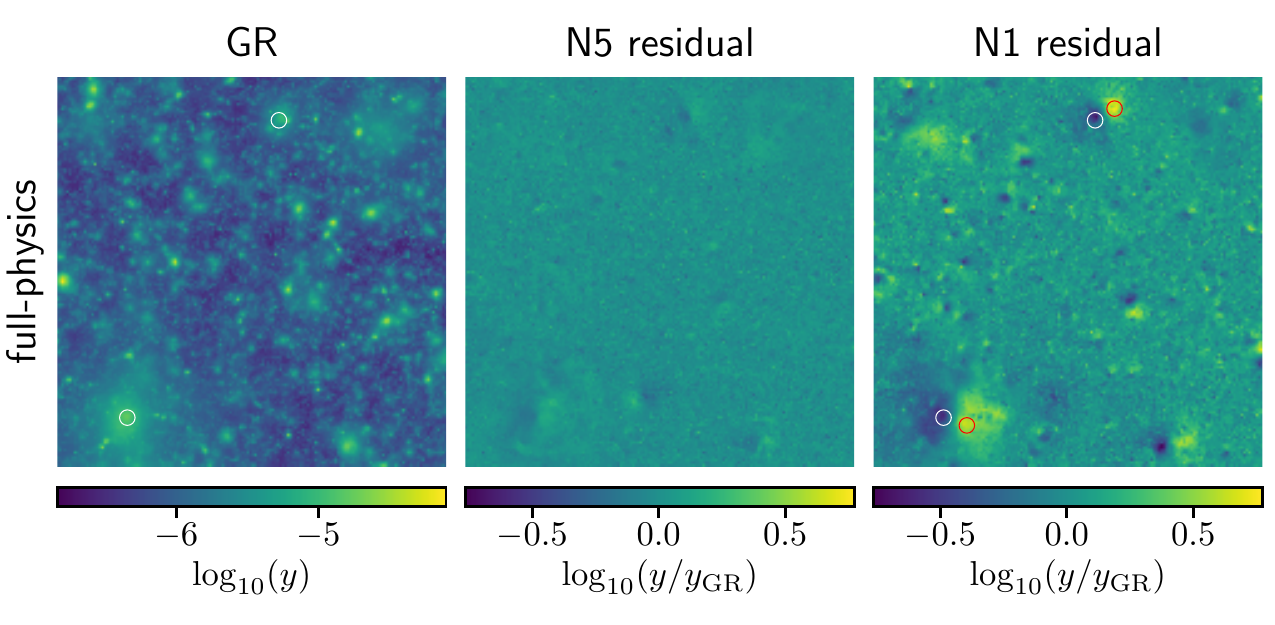}
\caption{(Colour Online) Map of the thermal SZ effect in GR (\textit{left column}) and its relative enhancement in N5 (\textit{middle column}) and N1 (\textit{right column}). The maps have a side-length of $1^\circ$ and a $512\times512$-pixel resolution, and have been constructed from the {\sc shybone} simulations (see Sec.~\ref{sec:methods}). The $y$-parameter is computed for each pixel using Eq.~(\ref{eq:pixel_ysz}). The rings indicate two haloes whose positions are shifted in N1 (\textit{red}) relative to GR (\textit{white}).}
\label{fig:nDGP_thermal_sz}
\end{figure*}

\subsection{SZ maps}
\label{sec:methods:maps}

In the generation of each SZ map, a light cone is first constructed using our simulation snapshots with $z\leq3$. We use a field of view of $1^{\circ}\times1^{\circ}$ for the light cone, which is aligned along a specified direction from an imaginary observer placed at the centre of the simulation box at $z=0$. The box is repeated along this direction and, at a given distance from the observer, the snapshot that is closest to the corresponding redshift is used. Each snapshot is randomly rotated and shifted in order to reduce statistical correlations caused by the repetition of the box.

The $1^{\circ}\times1^{\circ}$ field of view is split into a $512\times 512$ grid of pixels, and an imaginary light ray is fired along the central axis of each pixel from $z=3$ to the observer. For each gas cell, an effective size, $s$, is defined, which can be used to determine whether it intersects with the light ray. By approximating the gas cells as spherical, the radius, $r_{\rm cell}$, of a gas cell can be estimated using:
\begin{equation}
    r_{\rm cell} = 2.5\left(\frac{3V_{\rm cell}}{4\pi}\right)^{\frac{1}{3}},
\end{equation}
where $V_{\rm cell}$ is the volume of the gas cell. This quantity is similar to the smoothing radius in smoothed-particle hydrodynamics, with the factor $2.5$ used to smooth the gas distribution. However, in the mock SZ map, the minimum length scale that is resolved (at a given distance) is the pixel side-length, $r_{\rm pixel}$. The effective size of a gas cell is therefore set as follows:
\begin{equation}
    s = 
    \begin{cases}
        r_{\rm pixel} & \text{if } r_{\rm cell} < r_{\rm pixel}.\\
        r_{\rm cell} & \text{if } r_{\rm cell} \geq r_{\rm pixel}.
    \end{cases}
\end{equation}
A gas cell contributes to the SZ signal of a pixel if the distance between its centre of mass and the light ray is smaller than $s$.

The tSZ effect is quantified by the Compton $y$-parameter, which can be computed via an integral of the electron pressure along the line of sight as follows:
\begin{equation}
    y = \frac{\sigma_{\rm T}}{m_{\rm e}c^2}\int n_{\rm e}T_{\rm gas}{\rm d}l,
    \label{eq:actual_ysz}
\end{equation}
where $\sigma_{\rm T}$ is the Thomson scattering cross section, $m_{\rm e}$ is the electron rest mass, $n_{\rm e}$ is the number density of free electrons and $T_{\rm gas}$ is the gas temperature. This is evaluated for each pixel $ij$ via a summation over all gas cells that intersect with the light ray:
\begin{equation}
    y_{ij} = \frac{\sigma_{\rm T}}{m_{\rm e}c^2}\sum_{\alpha}p_{\alpha}w_{\alpha,ij},
    \label{eq:pixel_ysz}
\end{equation}
where $w_{\alpha,ij}$ is a normalised smoothing kernel. The quantity $p_{\alpha}$ is given by:
\begin{equation}
    p_{\alpha} = \frac{N_{\rm e,\alpha}}{s_{\alpha}^2}T_{\alpha},
\end{equation}
where $N_{\rm e,\alpha}$, $s_{\alpha}$ and $T_{\alpha}$ are the electron number count, the effective size and the temperature of gas cell $\alpha$, respectively. Note that we have not accounted for the relativistic SZ (rSZ) effect in our calculations. The rSZ effect can induce a significant bias in the measurement of the $y$-parameter for the most massive clusters \citep[see, e.g.,][]{Erler:2017dok}. However, the effect is much smaller for lower-mass objects which have a lower gas temperature. Since our simulations contain only galaxy groups and low-mass clusters ($M_{500}\lesssim10^{14.5}M_{\odot}$), we expect that including the rSZ effect would have a modest impact on our tSZ power spectrum results. In particular, we expect the effect on the model differences to be very small, but this is something that should be tested in the future with large simulations that contain a fair sample of cluster-sized objects.

The kSZ effect is quantified by the $b$-parameter:
\begin{equation}
    b = \sigma_{\rm T}\int \frac{n_{\rm e}v_{\rm r}}{c}{\rm d}l,
    \label{eq:actual_bsz}
\end{equation}
where $v_{\rm r}$ is the radial component of the gas peculiar velocity and $b$ is positive (negative) for gas that is moving away from (towards) the observer. The $b$-parameter is equivalent to the CMB temperature fluctuation due to the kSZ effect: $b = -\Delta T/T$. This is evaluated for each pixel as follows:
\begin{equation}
    b_{ij} = \frac{\sigma_{\rm T}}{c}\sum_{\alpha}q_{\alpha}w_{\alpha,ij}.
    \label{eq:pixel_bsz}
\end{equation}
The quantity $q_{\alpha}$ is given by:
\begin{equation}
    q_{\alpha} = \frac{N_{\rm e,\alpha}}{s_{\alpha}^2}v_{\rm r,\alpha},
\end{equation}
where $v_{\rm r,\alpha}$ is the radial component of the peculiar velocity of gas cell $\alpha$.

We have generated 14 independent light cones, each aligned along a unique direction. The same set of directions has been used to construct the maps for each gravity model and for both the full-physics and non-radiative data. This means that for any two maps aligned in the same direction, the only differences are caused by the contrasting gravity models and hydrodynamics schemes. The tSZ and kSZ maps corresponding to one of the light cones, generated using the $f(R)$ simulations, are shown in Figs.~\ref{fig:thermal_sz} and \ref{fig:kinetic_sz}, respectively. For each figure, the GR maps are shown in the left column, with the map from the full-physics run in the top row and the map for the non-radiative simulation in the bottom row.

The bright yellow peaks in the tSZ maps, which correspond to a high $y$-parameter, trace hot gas within groups and clusters of galaxies. These peaks are found at the same positions in both the full-physics and non-radiative maps. However, the addition of feedback mechanisms, which create winds that heat up and blow gas out of haloes, cause the peaks to appear more diffuse in the full-physics map. The kSZ map is made up of dark and bright regions, which correspond to negative and positive values of the $b$-parameter, respectively.

Rather than the absolute maps of F6 and F5, which are visually very similar to the GR maps, we display residual maps to indicate the main differences. These are shown in the second and third columns of Figs.~\ref{fig:thermal_sz} and \ref{fig:kinetic_sz}. The tSZ residuals represent the enhancement of the $f(R)$ $y$-parameter with respect to GR for each pixel. The F6 residuals are quite close to zero across the field of view, owing to the efficient screening of the fifth force in galaxy groups and clusters for this model. However, for the F5 model, for which the fifth force is more prominent, the residuals appear more complex. Pairs of bright and dark regions, two of which are indicated by rings placed in Fig.~\ref{fig:thermal_sz}, are visible throughout the images. These are caused by the shift of halo positions in F5 compared to GR, with each dark (bright) region corresponding to the position in GR (F5). While this in itself does not provide useful information about the effect of the fifth force on the tSZ effect, we note that at the extremes the positive residuals ($\log_{10}(y/y_{\rm GR})\approx1.5$) are greater in magnitude than the negative residuals ($\log_{10}(y/y_{\rm GR})\approx-0.8$), indicating that the tSZ effect is strengthened on average in F5 compared to GR.

For the kSZ signal, the $f(R)$ gravity residuals correspond to the enhancement of the absolute value of the $b$-parameter with respect to GR. A higher value of $b$ indicates that gas is moving faster with respect to the CMB rest-frame. Many individual pixels gain much higher and much lower values of $b$, seemingly at random, across the field of view. This is caused by the effect of the fifth force on the motion of the gas. Pairs of bright and dark regions are also just visible in the F5 residual map, again corresponding to the relative shifts in halo position with respect to GR.

In the rightmost columns of Figs.~\ref{fig:thermal_sz} and \ref{fig:kinetic_sz}, we show the relative difference between the full-physics and non-radiative GR maps. The tSZ results indicate that within haloes the tSZ signal is suppressed (dark blue regions) by up to 86\% and boosted outside haloes (bright yellow regions) by up to 173\%. This is caused by the ejection of gas from haloes by feedback mechanisms, causing the electron pressure to be lowered within haloes and raised outside haloes. For the kSZ results, as for the $f(R)$ gravity residuals, the value of $b$ is increased and reduced seemingly at random, owing to the unpredictable effects of the full-physics processes on the motion of the gas.

The nDGP tSZ maps for the same light cone are shown in Fig.~\ref{fig:nDGP_thermal_sz}, where recall that we do not have non-radiative runs. Again, the fifth force causes a shift in halo positions with respect to GR, and this is clearly visible for both nDGP models. The effect is greater in the N1 model, which is a stronger modification of GR than N5. We do not show the kSZ maps for nDGP, since these appear similar to the $f(R)$ maps and do not offer extra information.

\begin{figure*}
\centering
\includegraphics[width=0.7\textwidth]{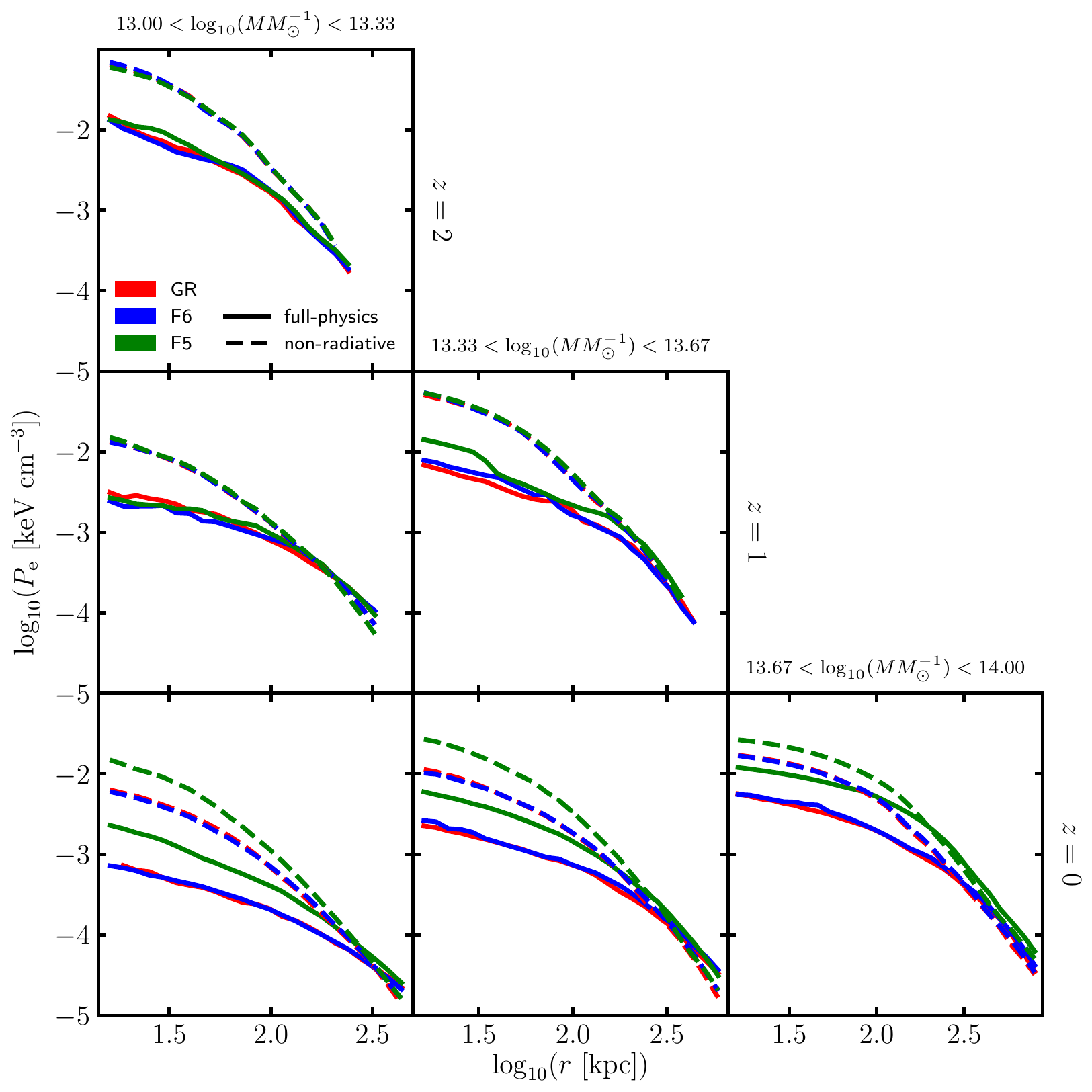}
\caption{(Colour Online) Stacked electron pressure profiles for haloes from three mass bins in the range $10^{13}M_{\odot} < M_{500} < 10^{14}M_{\odot}$ and redshifts 0, 1 and 2. The haloes have been identified from the {\sc shybone} simulations (see Sec.~\ref{sec:methods}), and have been generated for the GR (\textit{red}), F6 (\textit{blue}) and F5 (\textit{green}) gravity models, and for both the full-physics (\textit{solid lines}) and non-radiative (\textit{dashed lines}) hydrodynamics schemes.}
\label{fig:pressure}
\end{figure*}

\begin{figure*}
\centering
\includegraphics[width=0.7\textwidth]{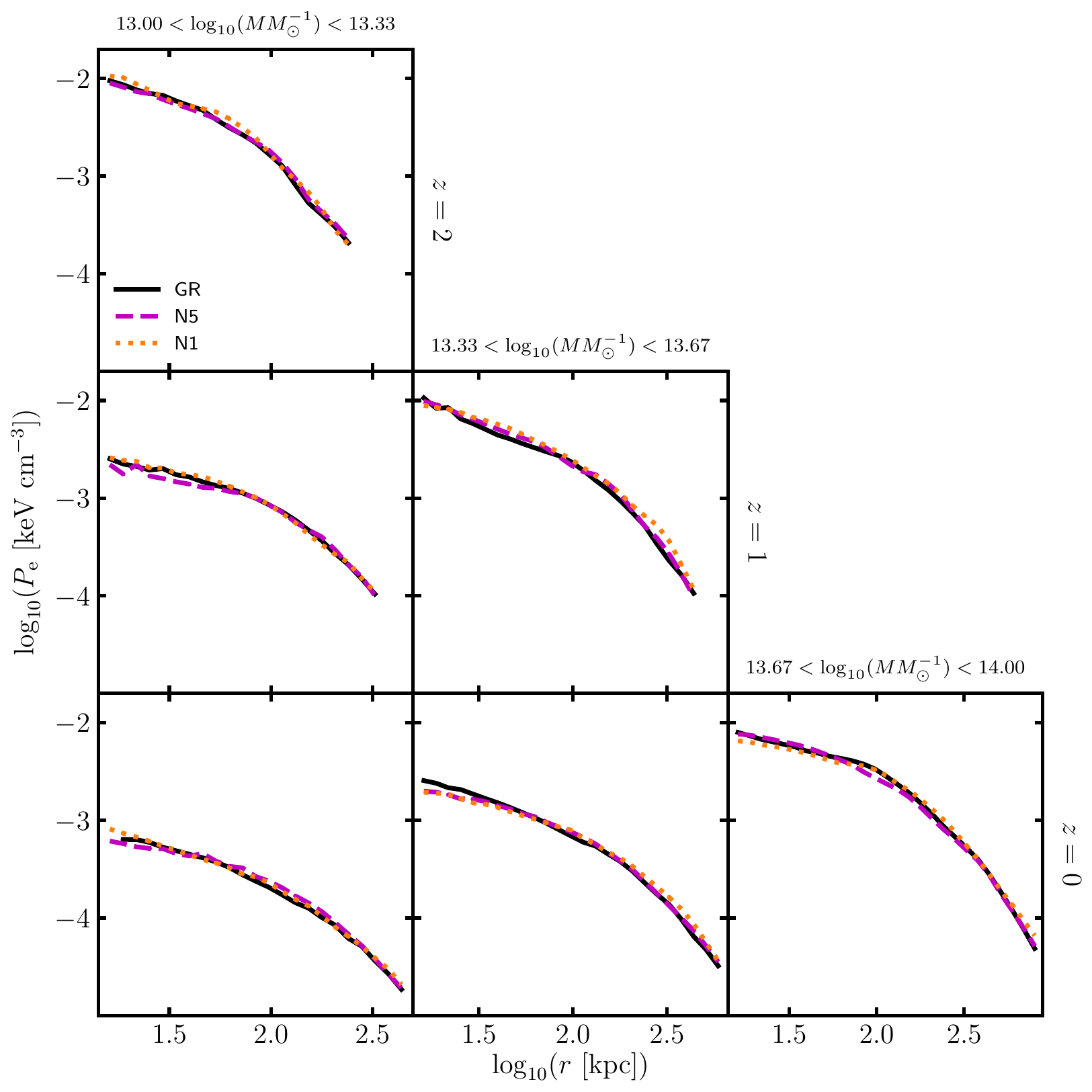}
\caption{(Colour Online) Stacked electron pressure profiles for haloes from three mass bins in the range $10^{13}M_{\odot} < M_{500} < 10^{14}M_{\odot}$ and redshifts 0, 1 and 2. The haloes have been identified from the full-physics {\sc shybone} simulations (see Sec.~\ref{sec:methods}), and have been generated for the GR (\textit{black solid}), N5 (\textit{magenta dashed}) and N1 (\textit{orange dotted}) gravity models.}
\label{fig:nDGP_pressure}
\end{figure*}

\begin{figure*}
\centering
\includegraphics[width=1.0\textwidth]{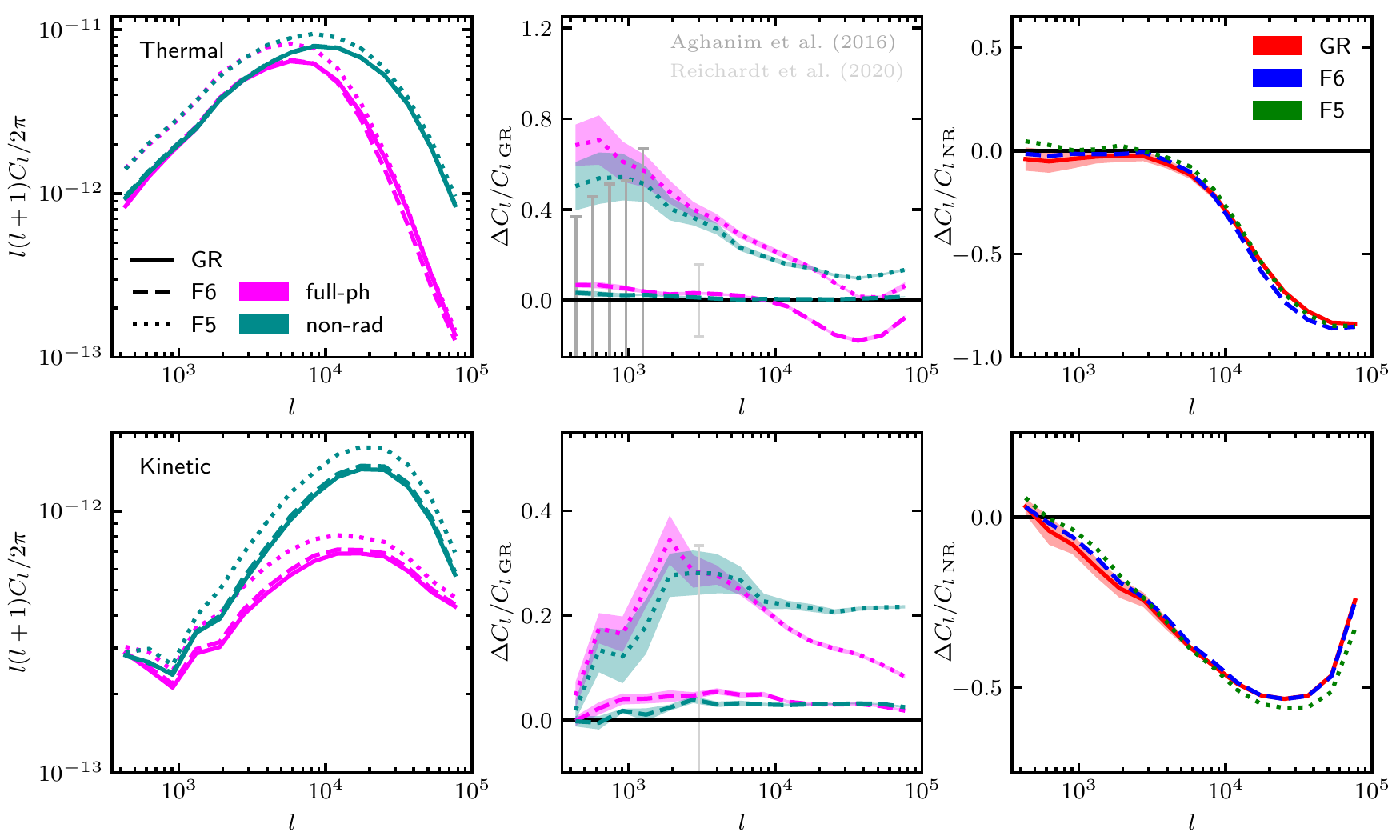}
\caption{(Colour Online) Angular power spectra and their relative differences plotted as a function of the angular wavenumber. The data has been generated from maps of the thermal (\textit{top row}) and kinetic (\textit{bottom row}) SZ signals, which have been created using the {\sc shybone} simulations (see Sec.~\ref{sec:methods}). \textit{Left column}: mean angular power spectrum plotted for GR (\textit{solid lines}), F6 (\textit{dashed lines}) and F5 (\textit{dotted lines}), including data from the full-physics (\textit{magenta}) and non-radiative (\textit{cyan}) simulations. \textit{Middle column}: mean relative enhancement of the F6 (\textit{dashed lines}) and F5 (\textit{dotted lines}) angular power spectra with respect to GR, plotted for the full-physics (\textit{magenta}) and non-radiative (\textit{cyan}) simulations. The standard error of the mean is indicated by the shaded regions. The error bars indicate the precision of the latest observations from the Planck \citep{Aghanim:2015eva} and South Pole Telescope \citep{Reichardt:2020jrr} collaborations. \textit{Right column}: mean relative enhancement of the full-physics angular power spectra with respect to the non-radiative data, plotted for GR (\textit{red}), F6 (\textit{blue}) and F5 (\textit{green}). For clarity, the standard error is shown for GR only.}
\label{fig:power_spectra}
\end{figure*}

\begin{figure*}
\centering
\includegraphics[width=0.67\textwidth]{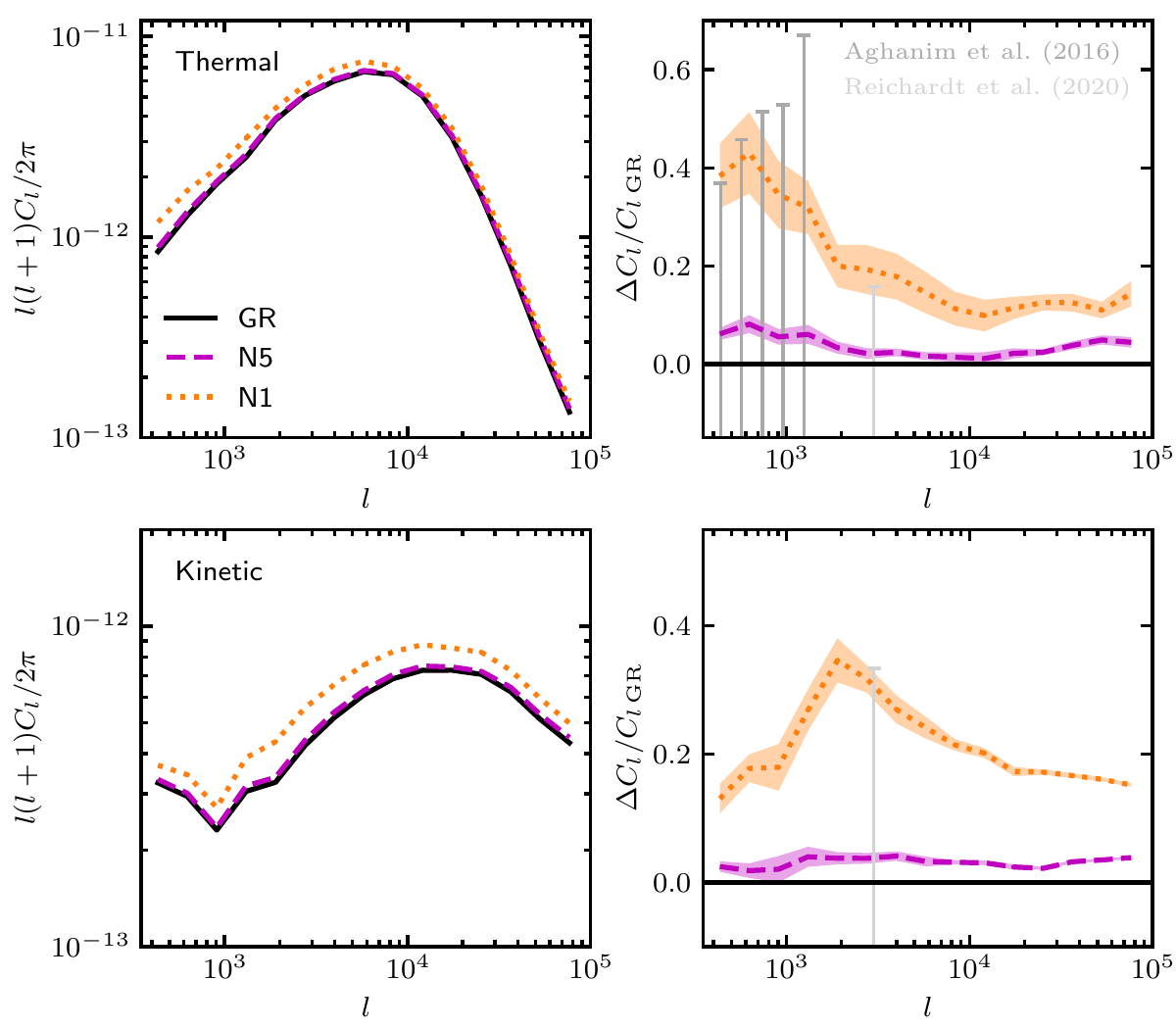}
\caption{Angular power spectra and their relative differences plotted as a function of the angular wavenumber. The data has been generated from maps of the thermal (\textit{top row}) and kinetic (\textit{bottom row}) SZ signals, which have been created using the {\sc shybone} simulations (see Sec.~\ref{sec:methods}). \textit{Left column}: mean angular power spectrum plotted for GR (\textit{solid lines}), N5 (\textit{dashed lines}) and N1 (\textit{dotted lines}). \textit{Right column}: mean relative enhancement of the N5 and N1 angular power spectra with respect to GR. The standard error of the mean is indicated by the shaded regions. The error bars indicate the precision of the latest observations from the Planck \citep{Aghanim:2015eva} and South Pole Telescope \citep{Reichardt:2020jrr} collaborations.}
\label{fig:nDGP_power_spectra}
\end{figure*}

\section{Results}
\label{sec:results}

This section gives the main results of this paper. In Sec.~\ref{sec:results:profiles}, we analyse the effects of baryonic processes and the fifth force on the stacked electron pressure profiles of FOF groups from our simulations. Then, in Sec.~\ref{sec:results:power}, we discuss the effects on the tSZ and kSZ angular power spectra. Finally, in Sec.~\ref{sec:results:transverse_momentum}, we examine the effects on the power spectrum of the transverse component of the electron momentum field, which is closely related to the kSZ angular power spectrum.

\subsection{Electron pressure profiles}
\label{sec:results:profiles}

We show the stacked electron pressure profiles at $z=0$, $z=1$ and $z=2$ in Fig.~\ref{fig:pressure} and Fig.~\ref{fig:nDGP_pressure} for $f(R)$ gravity and nDGP, respectively. Three equally spaced logarithmic mass bins, which span the range $13 < \log_{10}(M_{500}M_{\odot}^{-1}) < 14$, are considered. The volume-weighted electron pressure is measured in radial bins for each halo. This is given by the following:
\begin{equation}
    \bar{P}_{\rm e} = \frac{\sum_iP_{{\rm e},i}V_i}{\sum_iV_i},
\end{equation}
where $P_{{\rm e},i}$ and $V_i$ are the electron pressure and volume of gas cell $i$, and the summations are evaluated over all gas cells whose centres of mass are within the current bin. The median profile is measured for each radial bin using the haloes enclosed in each mass bin, and is displayed in the figures. Because of the limitations in the size of the halo population at higher masses, only the lowest-mass bin is shown at $z=2$ and the highest-mass bin is 
not shown for $z=1$. For the highest-mass bin shown at each redshift, because the halo number is relatively small, some haloes are also excluded from each model to ensure that the same halo population is used in all models. Any small difference in population could otherwise have a significant effect in these bins, which contain only $\sim10$ haloes each. This consideration is not required for the other bins, which have $\gtrsim25$ haloes each. We are also unable to include data at $M_{500}>10^{14}M_{\odot}$, for which there are only a few haloes for each model.

For haloes in F5 at sufficiently low redshift, we find that the fifth force boosts the electron pressure. This is caused by the increase in the temperature of the intra-cluster gas, which results from the deepened gravitational potential well \citep[e.g.,][]{He:2015mva,Mitchell:2020aep}. This indicates that the tSZ signal from individual haloes is expected to be significantly enhanced in F5. The magnitude of the background scalar field, $|f_R|$, increases with time, and as a result the chameleon mechanism is more efficient at screening the fifth force at earlier times. This explains why the enhancement of the pressure in F5 vanishes for $z\gtrsim1$. On the other hand, the background scalar field in the F6 model is $10$ times weaker than in F5, and as a result the fifth force is efficiently screened within group- and cluster-sized haloes even at $z=0$.

On the other hand, Fig.~\ref{fig:nDGP_pressure} shows much smaller differences in the electron pressure profiles in GR and nDGP than in Fig.~\ref{fig:pressure}, especially at lower $z$ ($z\lesssim1$). This is because the Vainshtein mechanism is much more efficient than the chameleon mechanism at screening out the fifth force within haloes at low redshifts --- for the latter, depending on the value of $|f_{R0}|$ in the two $f(R)$ models studied here, group-sized objects could be partially or completely unscreened at low $z$, while for the former the screening efficiency is similar for haloes of different masses \citep[see, e.g., Fig.~8 of][]{Hernandez-Aguayo:2020kgq}, including the ones as small as $\sim10^{11.7}h^{-1}M_\odot$, with the fifth force always being strongly suppressed in the inner regions of haloes, at all redshifts.

By comparing the non-radiative and full-physics data in Fig.~\ref{fig:pressure}, we can see that the additional baryonic processes that are present in the latter act to suppress the pressure at the inner halo regions. This can be caused by, for example, the blowing out of gas by black hole feedback which lowers the density of electrons. Note that, while the electron pressure profiles differ significantly between the full-physics and non-radiative runs, the relative enhancement of F5 with respect to GR seems to be consistent in both cases.

\subsection{tSZ and kSZ power spectra}
\label{sec:results:power}

We have used our SZ maps (see Sec.~\ref{sec:methods:maps}) to generate the tSZ and kSZ angular power spectra for the $f(R)$ and nDGP models. The power has been measured for each of the 14 maps in bins of the angular wavenumber $l$. For each bin, the mean power and the mean relative difference in the power between gravity models and hydrodynamics schemes have been measured.

From the $f(R)$ gravity results, shown in Fig.~\ref{fig:power_spectra}, we find that the fifth force and the extra baryonic processes that are found in the full-physics simulations have different effects: the middle column shows that, for the non-radiative data, the tSZ and kSZ power spectra are both enhanced in $f(R)$ gravity relative to GR; and the right column shows that the power is suppressed in the full-physics runs relative to the non-radiative runs, particularly at smaller scales. The latter is consistent with literature:  \citet{McCarthy:2013qva} showed that, at scales $l\gtrsim1000$, the tSZ power is suppressed by the ejection of gas by black hole feedback; and \citet{Park:2017amo} found that the kSZ power is suppressed by both the locking away of free electrons in stars, black holes and neutral gas (at all scales), and the ejection of gas through black hole feedback (at smaller scales). For our data, this suppression by baryonic processes occurs at $l\gtrsim3000$ for the tSZ power and at $l\gtrsim500$ for the kSZ power. The shape and amplitude of this suppression is very similar for each gravity model, as shown in the right column: the tSZ power is suppressed by up to $\sim85\%$ and the kSZ power is suppressed by up to $\sim50\%$.

With the extra baryonic processes of cooling, star formation and feedback absent, the tSZ power is enhanced by the fifth force on all scales. The enhancement is greater in F5 than in F6, with peaks of $\sim50\%$ and a few percent, respectively, at $l<1000$. However, due to the relatively small size of the fields of view in our light cones, we cannot measure the angular power spectra at $l\lesssim500$, and so it is unclear what the asymptotic behaviour at large angular scales is, for which future works with larger simulations are needed. For the kSZ power, a roughly constant enhancement is observed (of $\sim22\%$ for F5 and $\sim3\%$ for F6) at scales $l\gtrsim3000$, with a downturn at larger scales ($l\simeq2000$). The presence of the fifth force speeds up the formation of large-scale structures, boosting the abundance and peculiar velocity of groups and clusters of galaxies and, in turn, the tSZ and kSZ power spectra. In addition to this, the electron pressure profiles of haloes at a given mass are also enhanced, as discussed in Sec.~\ref{sec:results:profiles}, which could further boost the tSZ signal and tSZ power spectrum at small angular scales (the relation between the latter and halo electron pressure profiles, however, is more complicated, as we will discuss toward the end of this subsection).

The enhancement of the kSZ power by $\sim22\%$ in F5 is higher than predicted by \citet{Bianchini:2015iaa} and \citet{Roncarelli:2018kud}, who estimated an enhancement of about $15\%$ for the same model using analytical predictions and hydrodynamical simulations, respectively. We remark that our results use only the redshift range $z\leq3$ while these works used redshifts up to 9.9 and 15, including the epoch of reionisation which can have a substantial contribution to the total kSZ power. The fifth force is expected to be screened for $z\gtrsim3$, which can explain why the kSZ signal (an integral over the redshift range) shows less deviation from GR in these works. Our smaller redshift range $z\leq3$ also explains why the amplitude of the kSZ power in Fig.~\ref{fig:power_spectra} is lower than is typically predicted in literature \citep[e.g.,][]{Park:2017amo}.

The {\sc shybone} simulations are the first to simultaneously compute the fifth force of $f(R)$ gravity while incorporating full baryonic physics. The interplay between these two competing mechanisms in the full-physics simulations is therefore of particular interest. According to the middle column of Fig.~\ref{fig:power_spectra}, the extra processes in the full-physics simulations have a non-negligible effect on the relative difference between $f(R)$ gravity and GR. For the tSZ power, a suppression of the $f(R)$ enhancement is observed at very small scales ($l\gtrsim10000$), such that the F5 power is brought close to the GR power, and the F6 power becomes $\sim20\%$ lower than GR. For the kSZ power, the F5 enhancement is again suppressed at these scales, while there appears to be little change for F6. 

We note that these results are likely to be sensitive to the choice of full-physics parameters implemented by {\sc shybone}. Given that feedback is not fully understood theoretically or from observations, there is a non-negligible uncertainty in the results at small scales. In order to avoid potentially biased results, constraints should instead be made using large scales where the details of baryonic processes are not as prominent. For the tSZ power, these scales correspond to $l\lesssim3000$, although we note that even this range could be sensitive to the full-physics parameters. Our simulations predict enhancements of $\sim40$-$70\%$ in F5 and less than $10\%$ in F6, relative to GR, at these scales. For the kSZ power spectrum, again star formation, feedback and cooling appear to have a non-negligible effect at all of the scales tested in this work. However, the model differences between $f(R)$ gravity and GR do not differ significantly in the non-radiative and full-physics simulations for scales $l\lesssim10^4$. In this scale range, we observe relative differences of up to $\sim35\%$ and $\sim5\%$ between GR and the F5 and F6 models, respectively. Note that the non-radiative runs could be considered as an extreme case of the hydrodynamics scheme, with the most interesting physical processes neglected, and for this reason we expect that slight variations of the baryonic model should produce milder differences from the IllustrisTNG model than what is observed between the full-physics and non-radiative curves here. Also note that, due to the small box size, the full-physics runs used in this work could suffer from significant sample variance, e.g., due to a few large haloes experiencing unusually strong feedback in one model and not another; again, having a large simulation box in the future will help to address this question.

The tSZ and kSZ power spectra for the nDGP model (for full-physics only), are shown in Fig.~\ref{fig:nDGP_power_spectra}. As for the $f(R)$ model, the fifth force of nDGP enhances the power on all probed scales: the tSZ power is enhanced by up to $\sim40\%$ in N1 and less than $10\%$ in N5; and the kSZ power is enhanced by up to $\sim35\%$ in N1 and $\sim5\%$ in N5. However, given the absence of a non-radiative simulation for the nDGP model, we note that it is possible that these differences could be sensitive to baryonic physics, as in $f(R)$ gravity. 

Interestingly, the tSZ power spectrum at high $l$ is significantly enhanced in N1, even though the pressure profiles (Fig.~\ref{fig:nDGP_pressure}) do not appear to show a clear deviation from GR. There are a few reasons why this can happen. First of all, the tSZ power receives contributions from outside haloes as well as from within. The fourth column of Fig.~\ref{fig:thermal_sz} indicates that outside haloes the tSZ signal can be boosted by the ejection of gas by feedback. The presence of the fifth force is expected to result in the feedback being triggered earlier, which can cause the power to be enhanced relative to GR at angular scales corresponding roughly to halo sizes\footnote{The fifth force also enhances matter clustering on large scales overall, and this is expected to be reflected in the clustering of free electrons.}. Secondly, smaller angular scales receive a greater contribution from higher redshifts \citep[see, e.g.,][]{McCarthy:2013qva}. In F5, the fifth force is efficiently screened for $z\gtrsim1$, but in N1 it can still reach a few percent of the strength of the Newtonian force at the radius $R_{200}$ and $\sim8\%$ outside it at $z\sim2$ \citep[e.g.,][]{Hernandez-Aguayo:2020kgq}. This means that the SZ power at high $l$ can be enhanced by a greater amount in N1 than in F5. In fact, for the mass bin shown at $z=2$, the N1 pressure profile is enhanced by $\sim9\%$ with respect to GR, and we also find that the nDGP pressure is enhanced in lower-mass bins which are not shown in Fig.~\ref{fig:nDGP_pressure}. We have further verified (though not shown here) that at $z\gtrsim1.5$, the 3D electron pressure power spectrum is significantly enhanced even at high $k$ values well inside the 1-halo regime.

The results discussed in this section indicate that the tSZ and kSZ power have the potential to effectively probe $f(R)$ gravity and nDGP at large scales. To demonstrate this, we have included error bars in Figs.~\ref{fig:power_spectra} and \ref{fig:nDGP_power_spectra} to indicate the uncertainties of the latest tSZ and kSZ observations from the Planck \citep{Aghanim:2015eva} and South Pole Telescope \citep[SPT,][]{Reichardt:2020jrr} collaborations. The $\sim16\%$ precision of the tSZ measurement by SPT is sufficient to distinguish the F5 model from GR at $l=3000$, while the Planck measurements have sufficient precision to distinguish F5 at large angular scales ($l\lesssim500$). The $33\%$ precision of the kSZ measurement by SPT has a similar magnitude to the relative enhancements of the F5 and N1 models with respect to GR, indicating that more precise measurements from future surveys will be capable of ruling out these models. However, in order to avoid bias from uncertain baryonic physics, it will be necessary to use a range of full-physics parameters to confirm that reliable constraints can be achieved at these angular scales. It will also be important to revisit this study using simulations with a greater box size that can accurately probe the tSZ power up to angular scales $l\sim100$, where the precision of the Planck measurements is particularly high \citep{Aghanim:2015eva}. Finally, understanding the degeneracies between MG and variations in other cosmological parameters is also critical in order to have unbiased constraints.

Before finishing this section, let us note that, despite the qualitative difference in their respective screening mechanisms -- Vainshtein screening is always efficient inside dark matter haloes while the same cannot be said about the chameleon mechanism (cf.~Figs.~\ref{fig:pressure} and \ref{fig:nDGP_pressure}) -- the enhancements of both the tSZ and kSZ power spectra are very similar in these two models.

\begin{figure*}
\centering
\includegraphics[width=1.0\textwidth]{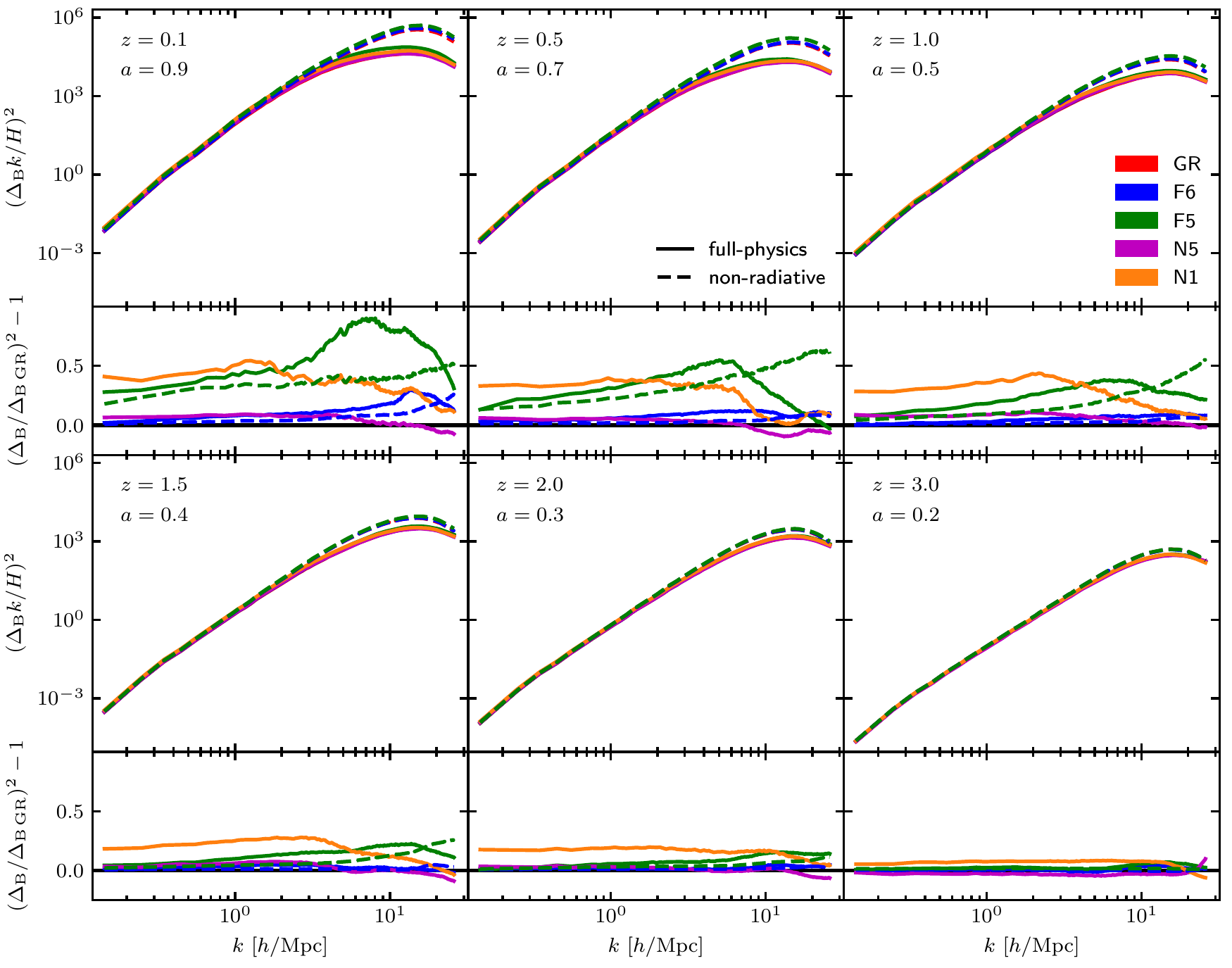}
\caption{(Colour Online) Power spectrum of the transverse component of the electron momentum plotted against the wavenumber. The data has been generated for six different redshifts (annotated) using the {\sc shybone} simulations (see Sec.~\ref{sec:methods:simulations}) for both the non-radiative (\textit{dashed lines}) and full-physics (\textit{solid lines}) hydrodynamics schemes. In addition to GR (\textit{red lines}), data is plotted for F6 (\textit{blue lines}), F5 (\textit{green lines}), N5 (\textit{magenta lines}) and N1 (\textit{orange lines}). The lower sub-panels show the relative enhancement of the MG (F6, F5, N5 and N1) power spectra with respect to GR.}
\label{fig:transverse_momentum}
\end{figure*}

\begin{figure}
\centering
\includegraphics[width=\columnwidth]{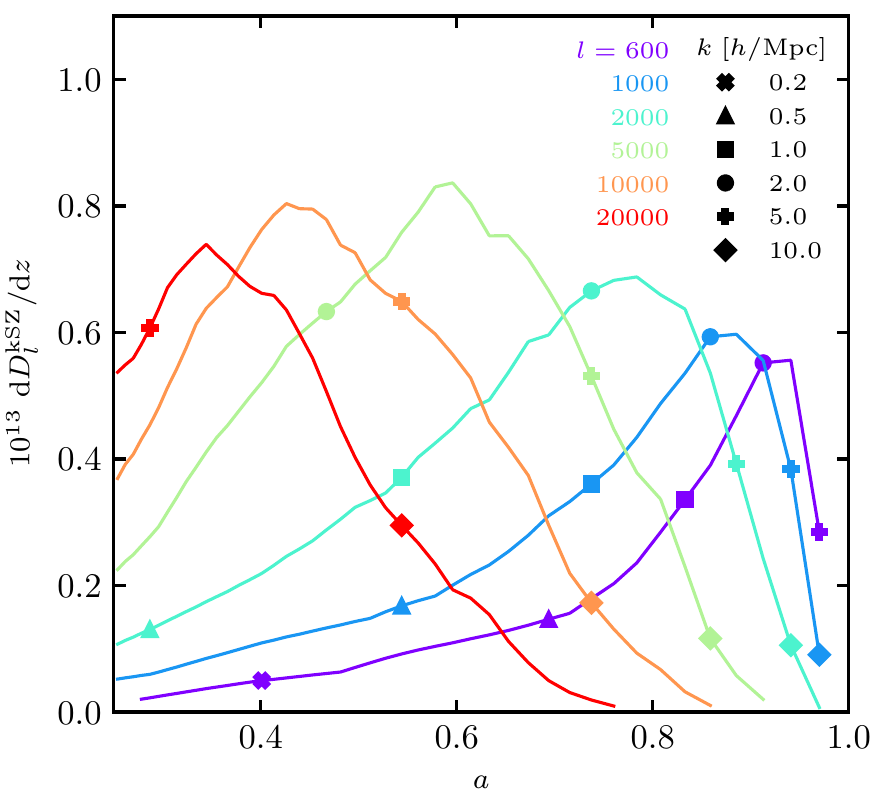}
\caption{(Colour Online) Derivative of the GR kSZ angular power spectrum as a function of the cosmological scale factor $a$ for six different values of the angular wavenumber $l$. The values are computed from the full-physics \textsc{shybone} simulation (see Sec.~\ref{sec:methods:simulations}) using the Limber approximation (Eq.~\ref{eq:limber}). The scales that are spanned by each $l$ value are indicated by markers which represent unique values of the wavenumber $k$.}
\label{fig:ksz_contribution}
\end{figure}

\subsection{Transverse momentum power spectrum}
\label{sec:results:transverse_momentum}

In order to understand the (similar) effects of $f(R)$ gravity and nDGP on the kSZ power in more detail, we have measured the power spectrum of the transverse component of the electron momentum field which, in the small-angle limit, can be related to the kSZ angular power spectrum using the Limber approximation \citep[e.g.,][]{2012ApJ...756...15S}:
\begin{equation}
\begin{aligned}
    & C_l^{\rm kSZ} = \frac{8\pi^2}{(2l+1)^3}\left(\frac{\sigma_{\rm T}\bar{\rho}_{\rm gas,0}}{\mu_{\rm e}m_{\rm p}}\right)^2 \\
    & \times \int_0^{z_{\rm re}}\frac{{\rm d}z}{c}(1+z)^4\chi^2\Delta_{\rm B}^2(k,z)e^{-2\tau(z)}\frac{x(z)}{H(z)},
\end{aligned}
\label{eq:limber}
\end{equation}
where $\bar{\rho}_{\rm gas,0}$ is the present-day mean background gas density, $\mu_{\rm e}m_{\rm p}$ is the mean gas mass per electron, $z_{\rm re}$ is the redshift at the epoch of reionisation, $\chi$ is the fraction of electrons that are ionised, $k=l/x$ is the wavenumber, $x(z)=\int_0^z(c{\rm d}z'/H(z'))$ is the comoving distance at redshift $z$, and the optical depth, $\tau$, is given by:
\begin{equation}
    \tau(z) = \sigma_{\rm T}c\int_0^z {\rm d}z'\frac{\bar{n}_{\rm e}(z')}{(1+z')H(z')}.
\end{equation}
We have computed the transverse momentum power, $\Delta_{\rm B}^2(k,z)$, using the electron momentum field $\boldsymbol{q}$ for a sample of snapshots from our simulations. This is defined $\boldsymbol{q} = \boldsymbol{v}(1 + \delta) = \boldsymbol{v}(n_{\rm e}/\bar{n}_{\rm e})$, where $\boldsymbol{v}$ is the velocity field of the gas. The power spectrum of the transverse momentum component, $\boldsymbol{q}_{\perp}$, is related to the power spectrum of the curl of the momentum field, $\boldsymbol{\nabla}\times\boldsymbol{q}$, by $P_{\boldsymbol{q_{\perp}}}=P_{\boldsymbol{\nabla}\times\boldsymbol{q}}/k^2$, and can be converted to the more commonly used definition $\Delta_{\rm B}^2 = P_{\boldsymbol{q_{\perp}}}k^3/(2\pi^2)$.

In Fig.~\ref{fig:transverse_momentum}, we show the dimensionless quantity $(\Delta_{\rm B}k/H)^2$ at six different redshifts for all gravity models and hydrodynamics schemes. In the lower sub-plots at each redshift, we show the MG enhancements of $\Delta_{\rm B}^2$ with respect to GR. For the non-radiative data, we see that the $f(R)$ enhancement is always increasing from large to small scales. The lowered enhancement at large scales is caused by the limited range of the fifth force, which is set by the Compton wavelength (Eq.~\eqref{eq:compton_wavelength}). For the full-physics data, the $f(R)$ enhancement follows a similar pattern at large scales but drops off for smaller scales ($k\gtrsim5h/{\rm Mpc}$) where baryonic processes are particularly prominent. For the nDGP data, the enhancement is roughly constant at large scales, since the fifth force in this model is long-range in the linear regime. There is again a suppression at small scales, which is likely to be caused by Vainshtein screening but could also be related to baryonic physics (we do not have non-radiative simulations to confirm the latter). For both $f(R)$ gravity and nDGP, the enhancement vanishes at higher redshifts where the amplitude of the scalar field is lower and the fifth force is screened out.

We also show, in Fig.~\ref{fig:ksz_contribution}, the derivative ${\rm d}D_l^{\rm kSZ}/{\rm d}z$, where 
\begin{equation}
D_l^{\rm kSZ} = l(l+1)C_l^{\rm kSZ}/(2\pi).
\end{equation}
This has been computed using the integrand and pre-factors in Eq.~(\ref{eq:limber}), and indicates the cosmic times and range of $k$-modes that have the greatest contribution to the kSZ angular power spectrum for different angular scales $l$. The enhancement of the kSZ power in F5 and N1 is observed to peak at $l=2000$ in Figs.~\ref{fig:power_spectra} and \ref{fig:nDGP_power_spectra}. From Fig.~\ref{fig:ksz_contribution}, we see that $D_{l=2000}^{\rm kSZ}$ receives a significant contribution from times $0.4 \lesssim a \lesssim 0.9$ and scales $1h/{\rm Mpc}\lesssim k \lesssim5h/{\rm Mpc}$. At these scales, the enhancement of $\Delta_{\rm B}^2$ peaks for N1 and has a similar magnitude for F5. The enhancements in these models span $\sim30\%$-$60\%$ over these scales and times, which is consistent with the peak enhancement of $C_{l}^{\rm kSZ}$. Going to larger angular scales ($600 < l < 1000$), $D_l^{\rm kSZ}$ is affected by lower $k$-modes (down to $\sim0.2h/{\rm Mpc}$) and lower redshifts. This then probes the larger scales (in Fig.~\ref{fig:transverse_momentum}) where the F5 fifth force is suppressed and the N1 enhancement levels off. This is consistent with Figs.~\ref{fig:power_spectra} and \ref{fig:nDGP_power_spectra}, where the kSZ power appears to be suppressed by a greater amount in F5 than in N1. Interestingly, this also implies that the enhancement of $C_l^{\rm kSZ}$ could be constant at angular scales larger than those available from our mock SZ maps. At smaller angular scales ($10000 \lesssim l \lesssim20000$), $D_l^{\rm kSZ}$ receives a significant contribution from high-$k$ modes ($2h/{\rm Mpc}\lesssim k\lesssim10h/{\rm Mpc}$), where Vainshtein screening suppresses the nDGP fifth force and the $f(R)$ fifth force is suppressed by baryonic processes (for the full-physics runs). In addition to this, $D_l^{\rm kSZ}$ is probing earlier times $0.25\leq a\lesssim 0.6$, where the scalar field amplitude is reduced in both models. This is therefore consistent with the lowered enhancement of $C_l^{\rm kSZ}$ at these angular scales. Note that the non-radiative $f(R)$ runs produce a higher kSZ power at high $l$ than the full-physics runs, which also agrees with the observation in Fig.~\ref{fig:transverse_momentum} that at large $k$ and $a\lesssim0.6$ the former has a larger transverse-momentum power spectrum.

The amplitude of $\Delta_{\rm B}k/H$ in Fig.~\ref{fig:transverse_momentum} appears to agree reasonably well with literature results \citep[e.g.,][]{Zhang:2003nr,2012ApJ...756...15S,Bianchini:2015iaa}, although it is slightly lower at large scales. We note that this is likely because the relatively small size $62h^{-1}{\rm Mpc}$ of our simulation box misses off longer-wavelength modes. It will therefore be useful to revisit this study with a larger box. The inclusion of longer-wavelength modes is expected to further suppress the F5 enhancement of $\Delta_{\rm B}^2$ at low-$k$ and to have little effect on the N1 enhancement.

We also note that for the entire $l$ range studied in Figs.~\ref{fig:power_spectra} and \ref{fig:nDGP_power_spectra} the kSZ power spectrum is dominated by modes with $k\gtrsim0.2h/{\rm Mpc}$ in the transverse-momentum power spectrum. From Fig.~\ref{fig:transverse_momentum}, we can see that in this regime galaxy formation has a non-negligible impact on $\Delta_{\rm B}$, which means that uncertainties in the subgrid physics can be an important theoretical systematic effect in using the kSZ power to test gravity models. Using kSZ data at $l<600$ may help reduce this effect, but the current simulation size does not allow a study of that range of $l$.

\section{Summary, Discussion and Conclusions}
\label{sec:conclusions}

Over the past couple of decades, great advances have been made in the measurement of the secondary anisotropies of the CMB caused by the SZ effect, including its thermal component and even its much smaller kinematic component. The angular power spectrum of the tSZ effect has been increasingly adopted as a probe of cosmological parameters that influence the growth of large-scale structures. Also, as observations of the kSZ power spectrum continue to improve, the latter has been identified as another potentially powerful probe of cosmology. The next generation of ground-based observatories \citep{Ade:2018sbj,Abazajian:2016yjj} look set to revolutionise the constraining power of these probes.

In this work, we have looked at the viability of using the angular power spectra of the tSZ and kSZ effects as large-scale probes of HS $f(R)$ gravity and nDGP, which are representative of a wide-range of MG theories which exhibit the chameleon and Vainshtein screening mechanisms, respectively. In order to do so, we have made use of the {\sc shybone} simulations (cf.~Sec.~\ref{sec:methods:simulations}), which are the first cosmological simulations that simultaneously incorporate full-physics plus HS $f(R)$ gravity \citep{Arnold:2019vpg} and nDGP \citep{Hernandez-Aguayo:2020kgq}. The simulations employ the IllustrisTNG full-physics model, which incorporates calibrated sub-resolution recipes for star formation and cooling as well as stellar and black hole feedback and allows realistic galaxy populations to be produced in hydrodynamical simulations.

Using these simulations, we have generated mock maps of the tSZ and kSZ signals (Sec.~\ref{sec:methods:maps}), and used these maps to measure the angular power spectra. Our results (Figs.~\ref{fig:power_spectra} and \ref{fig:nDGP_power_spectra}) indicate that the fifth force, present in $f(R)$ gravity and nDGP, and the subgrid baryonic physics have different effects on the tSZ and kSZ power spectra. The former enhances the power on all scales probed by our maps ($500\lesssim{l}\lesssim8\times10^4$) by boosting the abundance and peculiar velocity of large-scale structures (e.g., dark matter haloes and free electrons inside them), while the latter brings about a suppression on scales $l\gtrsim3000$ for the tSZ effect and on all tested scales for the kSZ effect. Even with both of these effects present, we find that the power can be significantly enhanced in $f(R)$ gravity and nDGP: by up to $60\%$ for the tSZ effect and $35\%$ for the kSZ effect for the F5 and N1 models; and by $5\%$-$10\%$ for F6 and N5, which correspond to relatively weak modifications of GR. In addition, we have computed the power spectrum of the transverse component of the electron momentum field (Sec.~\ref{sec:results:transverse_momentum}), which is closely related to the kSZ angular power spectrum. In particular, we show in Fig.~\ref{fig:ksz_contribution} that at angular sizes $l\geq600$ the kSZ signal is dominantly contributed by $k$-modes in the transverse-momentum power spectrum which are in the non-linear regime, and which are affected strongly by MG. The $k$-modes in the linear regime may contribute more to smaller $l$, but at least for $f(R)$ gravity the impact of MG at those $l$ values will be much less significant due to the finite range of the fifth force, as we can already see in Fig.~\ref{fig:power_spectra}.

We find that the relative difference between the MG models and GR is significantly affected by the additional baryonic processes that act in the full-physics simulations. Given that these processes are still relatively less well-constrained, this adds to the uncertainty in our theoretical predictions of the kSZ angular power spectra on small angular scales, e.g., $l>600$. Therefore, further work should be carried out using a range of full-physics parameters to precisely identify the scales on which constraints can be reliably made before the tSZ and kSZ power are used to probe $f(R)$ gravity and nDGP.

Finally, we note that the reason we are unable to study larger scales is the relatively small box size of the \textsc{shybone} simulations. We are preparing to run larger simulations with a re-calibrated full-physics model, and will redo this analysis in a future work.

\section*{Acknowledgements}

MAM is supported by a PhD Studentship with the Durham Centre for Doctoral Training in Data Intensive Science, funded by the UK Science and Technology Facilities Council (STFC, ST/P006744/1) and Durham University. CA and BL are supported by the European Research Council via grant ERC-StG-716532-PUNCA. BL is additionally supported by STFC Consolidated Grants ST/T000244/1 and ST/P000541/1. CH-A is supported by the Excellence Cluster ORIGINS which is funded by the Deutsche Forschungsgemeinschaft (DFG, German Research Foundation) under Germany's Excellence Strategy - EXC-2094 - 390783311. This work used the DiRAC@Durham facility managed by the Institute for Computational Cosmology on behalf of the STFC DiRAC HPC Facility (\url{www.dirac.ac.uk}). The equipment was funded by BEIS capital funding via STFC capital grants ST/K00042X/1, ST/P002293/1, ST/R002371/1 and ST/S002502/1, Durham University and STFC operations grant ST/R000832/1. DiRAC is part of the National e-Infrastructure.

\section*{Data availability}

The simulation data and results of this paper may be available upon request.




\bibliographystyle{mnras}
\bibliography{references} 

\begin{thebibliography}{}
\makeatletter
\relax
\def\mn@urlcharsother{\let\do\@makeother \do\$\do\&\do\#\do\^\do\_\do\%\do\~}
\def\mn@doi{\begingroup\mn@urlcharsother \@ifnextchar [ {\mn@doi@}
  {\mn@doi@[]}}
\def\mn@doi@[#1]#2{\def\@tempa{#1}\ifx\@tempa\@empty \href
  {http://dx.doi.org/#2} {doi:#2}\else \href {http://dx.doi.org/#2} {#1}\fi
  \endgroup}
\def\mn@eprint#1#2{\mn@eprint@#1:#2::\@nil}
\def\mn@eprint@arXiv#1{\href {http://arxiv.org/abs/#1} {{\tt arXiv:#1}}}
\def\mn@eprint@dblp#1{\href {http://dblp.uni-trier.de/rec/bibtex/#1.xml}
  {dblp:#1}}
\def\mn@eprint@#1:#2:#3:#4\@nil{\def\@tempa {#1}\def\@tempb {#2}\def\@tempc
  {#3}\ifx \@tempc \@empty \let \@tempc \@tempb \let \@tempb \@tempa \fi \ifx
  \@tempb \@empty \def\@tempb {arXiv}\fi \@ifundefined
  {mn@eprint@\@tempb}{\@tempb:\@tempc}{\expandafter \expandafter \csname
  mn@eprint@\@tempb\endcsname \expandafter{\@tempc}}}

\bibitem[\protect\citeauthoryear{Abazajian et~al.}{Abazajian
  et~al.}{2016}]{Abazajian:2016yjj}
Abazajian K.~N.,  et~al., 2016, preprint (\mn@eprint {arXiv} {1610.02743})

\bibitem[\protect\citeauthoryear{Ade et~al.}{Ade et~al.}{2019}]{Ade:2018sbj}
Ade P.,  et~al., 2019, \mn@doi [J. Cosmo. Astropart. Phys.]
  {10.1088/1475-7516/2019/02/056}, 02, 056

\bibitem[\protect\citeauthoryear{Aghanim et~al.}{Aghanim
  et~al.}{2016}]{Aghanim:2015eva}
Aghanim N.,  et~al., 2016, \mn@doi [Astron. Astrophys.]
  {10.1051/0004-6361/201525826}, 594, A22

\bibitem[\protect\citeauthoryear{Arnalte-Mur, Hellwing  \& Norberg}{Arnalte-Mur
  et~al.}{2017}]{Arnalte-Mur:2016alq}
Arnalte-Mur P.,  Hellwing W.~A.,   Norberg P.,  2017, \mn@doi [Mon. Not. Roy.
  Astron. Soc.] {10.1093/mnras/stx196}, 467, 1569

\bibitem[\protect\citeauthoryear{Arnold, Leo  \& Li}{Arnold
  et~al.}{2019}]{Arnold:2019vpg}
Arnold C.,  Leo M.,   Li B.,  2019, \mn@doi [Nature Astron.]
  {10.1038/s41550-019-0823-y}, 3, 945

\bibitem[\protect\citeauthoryear{Barreira, S\'anchez  \& Schmidt}{Barreira
  et~al.}{2016}]{PhysRevD.94.084022}
Barreira A.,  S\'anchez A.~G.,   Schmidt F.,  2016, \mn@doi [Phys. Rev.]
  {10.1103/PhysRevD.94.084022}, D94, 084022

\bibitem[\protect\citeauthoryear{Bianchini \& Silvestri}{Bianchini \&
  Silvestri}{2016}]{Bianchini:2015iaa}
Bianchini F.,  Silvestri A.,  2016, \mn@doi [Phys. Rev.]
  {10.1103/PhysRevD.93.064026}, D93, 064026

\bibitem[\protect\citeauthoryear{Bolliet, Comis, Komatsu  \&
  Mac\'\i{}as-P\'erez}{Bolliet et~al.}{2018}]{Bolliet:2017lha}
Bolliet B.,  Comis B.,  Komatsu E.,   Mac\'\i{}as-P\'erez J.~F.,  2018, \mn@doi
  [Mon. Not. Roy. Astron. Soc.] {10.1093/mnras/sty823}, 477, 4957

\bibitem[\protect\citeauthoryear{Bose \& Koyama}{Bose \&
  Koyama}{2017}]{Bose:2017dtl}
Bose B.,  Koyama K.,  2017, \mn@doi [J. Cosmo. Astropart. Phys.]
  {10.1088/1475-7516/2017/08/029}, 08, 029

\bibitem[\protect\citeauthoryear{Capozziello \& De~Laurentis}{Capozziello \&
  De~Laurentis}{2012}]{Capozziello:2012ie}
Capozziello S.,  De~Laurentis M.,  2012, \mn@doi [Annalen Phys.]
  {10.1002/andp.201200109}, 524, 545

\bibitem[\protect\citeauthoryear{Cataneo et~al.,}{Cataneo
  et~al.}{2015}]{PhysRevD.92.044009}
Cataneo M.,  et~al., 2015, \mn@doi [Phys. Rev.] {10.1103/PhysRevD.92.044009},
  D92, 044009

\bibitem[\protect\citeauthoryear{Cautun, Paillas, Cai, Bose, Armijo, Li  \&
  Padilla}{Cautun et~al.}{2018}]{Cautun:2017tkc}
Cautun M.,  Paillas E.,  Cai Y.-C.,  Bose S.,  Armijo J.,  Li B.,   Padilla N.,
   2018, \mn@doi [Mon. Not. Roy. Astron. Soc.] {10.1093/mnras/sty463}, 476,
  3195

\bibitem[\protect\citeauthoryear{De~Felice \& Tsujikawa}{De~Felice \&
  Tsujikawa}{2010}]{DeFelice:2010aj}
De~Felice A.,  Tsujikawa S.,  2010, \mn@doi [Living Rev. Rel.]
  {10.12942/lrr-2010-3}, 13, 3

\bibitem[\protect\citeauthoryear{De~Martino}{De~Martino}{2016}]{deMartino:2016xso}
De~Martino I.,  2016, \mn@doi [Phys. Rev] {10.1103/PhysRevD.93.124043}, D93,
  124043

\bibitem[\protect\citeauthoryear{De~Martino, De~Laurentis, Atrio-Barandela  \&
  Capozziello}{De~Martino et~al.}{2014}]{DeMartino:2013zua}
De~Martino I.,  De~Laurentis M.,  Atrio-Barandela F.,   Capozziello S.,  2014,
  \mn@doi [Mon. Not. Roy. Astron. Soc.] {10.1093/mnras/stu903}, 442, 921

\bibitem[\protect\citeauthoryear{Dvali, Gabadadze  \& Porrati}{Dvali
  et~al.}{2000}]{DVALI2000208}
Dvali G.,  Gabadadze G.,   Porrati M.,  2000, \mn@doi [Physics Letters B]
  {https://doi.org/10.1016/S0370-2693(00)00669-9}, 485, 208

\bibitem[\protect\citeauthoryear{Erler, Basu, Chluba  \& Bertoldi}{Erler
  et~al.}{2018}]{Erler:2017dok}
Erler J.,  Basu K.,  Chluba J.,   Bertoldi F.,  2018, \mn@doi [Mon. Not. Roy.
  Astron. Soc.] {10.1093/mnras/sty327}, 476, 3360

\bibitem[\protect\citeauthoryear{Falck, Koyama, Zhao  \& Cautun}{Falck
  et~al.}{2018}]{Falck:2017rvl}
Falck B.,  Koyama K.,  Zhao G.-B.,   Cautun M.,  2018, \mn@doi [Mon. Not. Roy.
  Astron. Soc.] {10.1093/mnras/stx3288}, 475, 3262

\bibitem[\protect\citeauthoryear{George et~al.}{George
  et~al.}{2015}]{George:2014oba}
George E.,  et~al., 2015, \mn@doi [Astrophys. J.]
  {10.1088/0004-637X/799/2/177}, 799, 177

\bibitem[\protect\citeauthoryear{Hammami \& Mota}{Hammami \&
  Mota}{2017}]{Hammami:2016npf}
Hammami A.,  Mota D.~F.,  2017, \mn@doi [Astron. Astrophys.]
  {10.1051/0004-6361/201629003}, 598, A132

\bibitem[\protect\citeauthoryear{He \& Li}{He \& Li}{2016}]{He:2015mva}
He J.-h.,  Li B.,  2016, \mn@doi [Phys. Rev.] {10.1103/PhysRevD.93.123512},
  D93, 123512

\bibitem[\protect\citeauthoryear{{He}, {Guzzo}, {Li}  \& {Baugh}}{{He}
  et~al.}{2018}]{2018NatAs...2..967H}
{He} J.-h.,  {Guzzo} L.,  {Li} B.,   {Baugh} C.~M.,  2018, \mn@doi [Nature
  Astron.] {10.1038/s41550-018-0573-2}, \href
  {https://ui.adsabs.harvard.edu/abs/2018NatAs...2..967H} {2, 967}

\bibitem[\protect\citeauthoryear{Hern\'andez-Aguayo, Hou, Li, Baugh  \&
  S\'anchez}{Hern\'andez-Aguayo et~al.}{2019}]{Hernandez-Aguayo:2018oxg}
Hern\'andez-Aguayo C.,  Hou J.,  Li B.,  Baugh C.~M.,   S\'anchez A.~G.,  2019,
  \mn@doi [Mon. Not. Roy. Astron. Soc.] {10.1093/mnras/stz516}, 485, 2194

\bibitem[\protect\citeauthoryear{Hern\'andez-Aguayo, Arnold, Li  \&
  Baugh}{Hern\'andez-Aguayo et~al.}{2020}]{Hernandez-Aguayo:2020kgq}
Hern\'andez-Aguayo C.,  Arnold C.,  Li B.,   Baugh C.~M.,  2020, preprint
  (\mn@eprint {arXiv} {2006.15467})

\bibitem[\protect\citeauthoryear{Horowitz \& Seljak}{Horowitz \&
  Seljak}{2017}]{Horowitz:2016dwk}
Horowitz B.,  Seljak U.,  2017, \mn@doi [Mon. Not. Roy. Astron. Soc.]
  {10.1093/mnras/stx766}, 469, 394

\bibitem[\protect\citeauthoryear{Hu \& Sawicki}{Hu \&
  Sawicki}{2007}]{Hu:2007nk}
Hu W.,  Sawicki I.,  2007, \mn@doi [Phys. Rev.] {10.1103/PhysRevD.76.064004},
  D76, 064004

\bibitem[\protect\citeauthoryear{Hurier \& Lacasa}{Hurier \&
  Lacasa}{2017}]{Hurier:2017jgi}
Hurier G.,  Lacasa F.,  2017, \mn@doi [Astron. Astrophys.]
  {10.1051/0004-6361/201630041}, 604, A71

\bibitem[\protect\citeauthoryear{Khoury \& Weltman}{Khoury \&
  Weltman}{2004a}]{Khoury:2003aq}
Khoury J.,  Weltman A.,  2004a, \mn@doi [Phys. Rev. Lett.]
  {10.1103/PhysRevLett.93.171104}, 93, 171104

\bibitem[\protect\citeauthoryear{Khoury \& Weltman}{Khoury \&
  Weltman}{2004b}]{Khoury:2003rn}
Khoury J.,  Weltman A.,  2004b, \mn@doi [Phys. Rev.]
  {10.1103/PhysRevD.69.044026}, D69, 044026

\bibitem[\protect\citeauthoryear{Koyama}{Koyama}{2016}]{Koyama:2015vza}
Koyama K.,  2016, \mn@doi [Rept. Prog. Phys.] {10.1088/0034-4885/79/4/046902},
  79, 046902

\bibitem[\protect\citeauthoryear{Koyama \& Silva}{Koyama \&
  Silva}{2007}]{PhysRevD.75.084040}
Koyama K.,  Silva F.~P.,  2007, \mn@doi [Phys. Rev.]
  {10.1103/PhysRevD.75.084040}, D75, 084040

\bibitem[\protect\citeauthoryear{Li, Zhao, Teyssier  \& Koyama}{Li
  et~al.}{2012}]{Li:2011vk}
Li B.,  Zhao G.-B.,  Teyssier R.,   Koyama K.,  2012, \mn@doi [JCAP]
  {10.1088/1475-7516/2012/01/051}, 01, 051

\bibitem[\protect\citeauthoryear{Li, He  \& Gao}{Li et~al.}{2016}]{Li:2015rva}
Li B.,  He J.-h.,   Gao L.,  2016, \mn@doi [Mon. Not. Roy. Astron. Soc.]
  {10.1093/mnras/stv2650}, 456, 146

\bibitem[\protect\citeauthoryear{Liu et~al.}{Liu et~al.}{2016}]{Liu:2016xes}
Liu X.,  et~al., 2016, \mn@doi [Phys. Rev. Lett.]
  {10.1103/PhysRevLett.117.051101}, 117, 051101

\bibitem[\protect\citeauthoryear{Ma \& Zhao}{Ma \& Zhao}{2014}]{Ma:2013taq}
Ma Y.-Z.,  Zhao G.-B.,  2014, \mn@doi [Phys. Lett. B]
  {10.1016/j.physletb.2014.06.066}, 735, 402

\bibitem[\protect\citeauthoryear{Marinacci et~al.}{Marinacci
  et~al.}{2018}]{Marinacci:2017wew}
Marinacci F.,  et~al., 2018, \mn@doi [Mon. Not. Roy. Astron. Soc.]
  {10.1093/mnras/sty2206}, 480, 5113

\bibitem[\protect\citeauthoryear{McCarthy, Brun, Schaye  \& Holder}{McCarthy
  et~al.}{2014}]{McCarthy:2013qva}
McCarthy I.~G.,  Brun A. M. C.~L.,  Schaye J.,   Holder G.~P.,  2014, \mn@doi
  [Mon. Not. Roy. Astron. Soc.] {10.1093/mnras/stu543}, 440, 3645

\bibitem[\protect\citeauthoryear{Mitchell, Arnold  \& Li}{Mitchell
  et~al.}{2020}]{Mitchell:2020aep}
Mitchell M.~A.,  Arnold C.,   Li B.,  2020, preprint (\mn@eprint {arXiv}
  {2010.11964})

\bibitem[\protect\citeauthoryear{Mota \& Shaw}{Mota \&
  Shaw}{2007}]{Mota:2006fz}
Mota D.~F.,  Shaw D.~J.,  2007, \mn@doi [Phys. Rev.]
  {10.1103/PhysRevD.75.063501}, D75, 063501

\bibitem[\protect\citeauthoryear{{Naiman} et~al.,}{{Naiman}
  et~al.}{2018}]{2018MNRAS.477.1206N}
{Naiman} J.~P.,  et~al., 2018, \mn@doi [\mnras] {10.1093/mnras/sty618}, \href
  {https://ui.adsabs.harvard.edu/abs/2018MNRAS.477.1206N} {477, 1206}

\bibitem[\protect\citeauthoryear{Nelson et~al.}{Nelson
  et~al.}{2018}]{Nelson:2017cxy}
Nelson D.,  et~al., 2018, \mn@doi [Mon. Not. Roy. Astron. Soc.]
  {10.1093/mnras/stx3040}, 475, 624

\bibitem[\protect\citeauthoryear{Paillas, Cautun, Li, Cai, Padilla, Armijo  \&
  Bose}{Paillas et~al.}{2019}]{10.1093/mnras/stz022}
Paillas E.,  Cautun M.,  Li B.,  Cai Y.-C.,  Padilla N.,  Armijo J.,   Bose S.,
   2019, \mn@doi [Mon. Not. Roy. Astron. Soc.] {10.1093/mnras/stz022}, 484,
  1149

\bibitem[\protect\citeauthoryear{{Pakmor} \& {Springel}}{{Pakmor} \&
  {Springel}}{2013}]{Pakmor2013}
{Pakmor} R.,  {Springel} V.,  2013, \mn@doi [\mnras] {10.1093/mnras/stt428},
  \href {https://ui.adsabs.harvard.edu/abs/2013MNRAS.432..176P} {432, 176}

\bibitem[\protect\citeauthoryear{{Pakmor}, {Bauer}  \& {Springel}}{{Pakmor}
  et~al.}{2011}]{Pakmor2011}
{Pakmor} R.,  {Bauer} A.,   {Springel} V.,  2011, \mn@doi [\mnras]
  {10.1111/j.1365-2966.2011.19591.x}, \href
  {https://ui.adsabs.harvard.edu/abs/2011MNRAS.418.1392P} {418, 1392}

\bibitem[\protect\citeauthoryear{Park, Alvarez  \& Bond}{Park
  et~al.}{2018}]{Park:2017amo}
Park H.,  Alvarez M.~A.,   Bond J.~R.,  2018, \mn@doi [Astrophys. J.]
  {10.3847/1538-4357/aaa0da}, 853, 121

\bibitem[\protect\citeauthoryear{Peirone, Raveri, Viel, Borgani  \&
  Ansoldi}{Peirone et~al.}{2017}]{Peirone:2016wca}
Peirone S.,  Raveri M.,  Viel M.,  Borgani S.,   Ansoldi S.,  2017, \mn@doi
  [Phys. Rev.] {10.1103/PhysRevD.95.023521}, D95, 023521

\bibitem[\protect\citeauthoryear{Pillepich et~al.}{Pillepich
  et~al.}{2018a}]{Pillepich:2017jle}
Pillepich A.,  et~al., 2018a, \mn@doi [Mon. Not. Roy. Astron. Soc.]
  {10.1093/mnras/stx2656}, 473, 4077

\bibitem[\protect\citeauthoryear{Pillepich et~al.}{Pillepich
  et~al.}{2018b}]{Pillepich:2017fcc}
Pillepich A.,  et~al., 2018b, \mn@doi [Mon. Not. Roy. Astron. Soc.]
  {10.1093/mnras/stx3112}, 475, 648

\bibitem[\protect\citeauthoryear{Reichardt et~al.}{Reichardt
  et~al.}{2020}]{Reichardt:2020jrr}
Reichardt C.,  et~al., 2020, preprint (\mn@eprint {arXiv} {2002.06197})

\bibitem[\protect\citeauthoryear{Roncarelli, Baldi  \&
  Villaescusa-Navarro}{Roncarelli et~al.}{2018}]{Roncarelli:2018kud}
Roncarelli M.,  Baldi M.,   Villaescusa-Navarro F.,  2018, \mn@doi [Mon. Not.
  Roy. Astron. Soc.] {10.1093/mnras/sty2225}, 481, 2497

\bibitem[\protect\citeauthoryear{Salvati, Douspis  \& Aghanim}{Salvati
  et~al.}{2018}]{Salvati:2017rsn}
Salvati L.,  Douspis M.,   Aghanim N.,  2018, \mn@doi [Astron. Astrophys.]
  {10.1051/0004-6361/201731990}, 614, A13

\bibitem[\protect\citeauthoryear{Schaye et~al.}{Schaye
  et~al.}{2015}]{Schaye:2014tpa}
Schaye J.,  et~al., 2015, \mn@doi [Mon. Not. Roy. Astron. Soc.]
  {10.1093/mnras/stu2058}, 446, 521

\bibitem[\protect\citeauthoryear{{Shaw}, {Rudd}  \& {Nagai}}{{Shaw}
  et~al.}{2012}]{2012ApJ...756...15S}
{Shaw} L.~D.,  {Rudd} D.~H.,   {Nagai} D.,  2012, \mn@doi [\apj]
  {10.1088/0004-637X/756/1/15}, \href
  {https://ui.adsabs.harvard.edu/abs/2012ApJ...756...15S} {756, 15}

\bibitem[\protect\citeauthoryear{Sievers et~al.}{Sievers
  et~al.}{2013}]{Sievers:2013ica}
Sievers J.~L.,  et~al., 2013, \mn@doi [J. Cosmo. Astropart. Phys.]
  {10.1088/1475-7516/2013/10/060}, 10, 060

\bibitem[\protect\citeauthoryear{Sotiriou \& Faraoni}{Sotiriou \&
  Faraoni}{2010}]{Sotiriou:2008rp}
Sotiriou T.~P.,  Faraoni V.,  2010, \mn@doi [Rev. Mod. Phys.]
  {10.1103/RevModPhys.82.451}, 82, 451

\bibitem[\protect\citeauthoryear{{Springel}}{{Springel}}{2010}]{2010MNRAS.401..791S}
{Springel} V.,  2010, \mn@doi [Mon. Not. Roy. Astron. Soc.]
  {10.1111/j.1365-2966.2009.15715.x}, \href
  {https://ui.adsabs.harvard.edu/\#abs/2010MNRAS.401..791S} {401, 791}

\bibitem[\protect\citeauthoryear{{Springel}, {White}, {Tormen}  \&
  {Kauffmann}}{{Springel} et~al.}{2001}]{springel2001}
{Springel} V.,  {White} S. D.~M.,  {Tormen} G.,   {Kauffmann} G.,  2001,
  \mn@doi [Mon. Not. Roy. Astron. Soc.] {10.1046/j.1365-8711.2001.04912.x},
  \href {https://ui.adsabs.harvard.edu/abs/2001MNRAS.328..726S} {328, 726}

\bibitem[\protect\citeauthoryear{Springel et~al.}{Springel
  et~al.}{2018}]{Springel:2017tpz}
Springel V.,  et~al., 2018, \mn@doi [Mon. Not. Roy. Astron. Soc.]
  {10.1093/mnras/stx3304}, 475, 676

\bibitem[\protect\citeauthoryear{{Sunyaev} \& {Zeldovich}}{{Sunyaev} \&
  {Zeldovich}}{1972}]{1972CoASP...4..173S}
{Sunyaev} R.~A.,  {Zeldovich} Y.~B.,  1972, Comments on Astrophysics and Space
  Physics, \href {https://ui.adsabs.harvard.edu/abs/1972CoASP...4..173S} {4,
  173}

\bibitem[\protect\citeauthoryear{{Sunyaev} \& {Zeldovich}}{{Sunyaev} \&
  {Zeldovich}}{1980}]{1980ARA&A..18..537S}
{Sunyaev} R.~A.,  {Zeldovich} I.~B.,  1980, \mn@doi [\araa]
  {10.1146/annurev.aa.18.090180.002541}, \href
  {https://ui.adsabs.harvard.edu/abs/1980ARA&A..18..537S} {18, 537}

\bibitem[\protect\citeauthoryear{Terukina, Lombriser, Yamamoto, Bacon, Koyama
  \& Nichol}{Terukina et~al.}{2014}]{Terukina:2013eqa}
Terukina A.,  Lombriser L.,  Yamamoto K.,  Bacon D.,  Koyama K.,   Nichol
  R.~C.,  2014, \mn@doi [J. Cosmo. Astropart. Phys.]
  {10.1088/1475-7516/2014/04/013}, 1404, 013

\bibitem[\protect\citeauthoryear{Terukina, Yamamoto, Okabe, Matsushita  \&
  Sasaki}{Terukina et~al.}{2015}]{Terukina:2015jua}
Terukina A.,  Yamamoto K.,  Okabe N.,  Matsushita K.,   Sasaki T.,  2015,
  \mn@doi [JCAP] {10.1088/1475-7516/2015/10/064}, 10, 064

\bibitem[\protect\citeauthoryear{Vainshtein}{Vainshtein}{1972}]{VAINSHTEIN1972393}
Vainshtein A.,  1972, \mn@doi [Physics Letters B]
  {https://doi.org/10.1016/0370-2693(72)90147-5}, 39, 393

\bibitem[\protect\citeauthoryear{{Vogelsberger}, {Genel}, {Sijacki}, {Torrey},
  {Springel}  \& {Hernquist}}{{Vogelsberger} et~al.}{2013}]{Vogelsberger2013}
{Vogelsberger} M.,  {Genel} S.,  {Sijacki} D.,  {Torrey} P.,  {Springel} V.,
  {Hernquist} L.,  2013, \mn@doi [\mnras] {10.1093/mnras/stt1789}, \href
  {https://ui.adsabs.harvard.edu/abs/2013MNRAS.436.3031V} {436, 3031}

\bibitem[\protect\citeauthoryear{{Weinberger} et~al.,}{{Weinberger}
  et~al.}{2017}]{2017MNRAS.465.3291W}
{Weinberger} R.,  et~al., 2017, \mn@doi [Mon. Not. Roy. Astron. Soc.]
  {10.1093/mnras/stw2944}, \href
  {https://ui.adsabs.harvard.edu/abs/2017MNRAS.465.3291W} {465, 3291}

\bibitem[\protect\citeauthoryear{Wilcox et~al.}{Wilcox
  et~al.}{2015}]{Wilcox:2015kna}
Wilcox H.,  et~al., 2015, \mn@doi [Mon. Not. Roy. Astron. Soc.]
  {10.1093/mnras/stv1366}, 452, 1171

\bibitem[\protect\citeauthoryear{Will}{Will}{2014}]{Will:2014kxa}
Will C.~M.,  2014, \mn@doi [Living Rev. Rel.] {10.12942/lrr-2014-4}, 17, 4

\bibitem[\protect\citeauthoryear{Zhang, Pen  \& Trac}{Zhang
  et~al.}{2004}]{Zhang:2003nr}
Zhang P.-J.,  Pen U.-L.,   Trac H.,  2004, \mn@doi [Mon. Not. Roy. Astron.
  Soc.] {10.1111/j.1365-2966.2004.07298.x}, 347, 1224

\makeatother
\end{thebibliography}





\bsp	
\label{lastpage}
\end{document}